\documentclass[9pt,sigconf]{acmart}
\usepackage{array}
\usepackage[english]{babel}
\usepackage[utf8]{inputenc}
\usepackage{blindtext}

\acmYear{2020}\copyrightyear{2020}
\setcopyright{rightsretained}
\acmConference[IMC '20]{ACM Internet Measurement Conference}{October 27--29, 2020}{Virtual Event, USA}
\acmBooktitle{ACM Internet Measurement Conference (IMC '20), October 27--29, 2020, Virtual Event, USA}
\acmPrice{}
\acmDOI{10.1145/3419394.3423658}
\acmISBN{978-1-4503-8138-3/20/10}



\usepackage{balance}

\usepackage{enumitem}
\setlist{nolistsep}
\usepackage{graphicx}
\usepackage{epstopdf}
\usepackage{grffile}
\usepackage{amsmath}
\usepackage{multicol,multirow}
\usepackage{rotating,url}
\usepackage{enumitem}
\usepackage{color}
\usepackage{subcaption}
\usepackage{xspace}
\usepackage{ifpdf}
\usepackage{mdwlist}
\usepackage{colortbl}
\usepackage{booktabs}
\usepackage{multirow}
\usepackage{enumitem}
\usepackage{xspace}
\usepackage{xfrac}
\usepackage[absolute,showboxes]{textpos}
\usepackage{afterpage}%
\newlength{\oldtextfloatsep}

\newcommand*\rot{\rotatebox{90}}

\newcommand{\ie}{i.e., \@}
\newcommand{\eg}{e.g., \@}

\newcommand{\etal}{et al.\xspace}
\newcommand{\covid}{COVID-19\xspace}
\newcommand{\one}{(1)\xspace}
\newcommand{\two}{(2)\xspace}
\newcommand{\three}{(3)\xspace}
\newcommand{\four}{(4)\xspace}

\newcommand{\isp}{ISP in Central Europe\xspace}
\newcommand{\nf}{NetFlow\xspace}
\newcommand{\ixpce}{IXP in Central Europe\xspace}
\newcommand{\ixpse}{IXP in Southern Europe\xspace}
\newcommand{\ixpus}{IXP at the US East Coast\xspace}
\newcommand{\ispS}{ISP-CE\xspace}
\newcommand{\ixpceS}{IXP-CE\xspace}
\newcommand{\ixpseS}{IXP-SE\xspace}
\newcommand{\ixpusS}{IXP-US\xspace}
\newcommand{\uniS}{EDU\xspace}

\newcommand{\afblock}[1]{\noindent{\textbf{#1 }}}

\usepackage{tabularx}
\newcolumntype{L}[1]{>{\raggedright\arraybackslash}p{#1}} %
\newcolumntype{C}[1]{>{\centering\arraybackslash}p{#1}} %
\newcolumntype{R}[1]{>{\raggedleft\arraybackslash}p{#1}} %

\newcommand{\todo}[1]{\textcolor{red}{\emph{#1}}}
\renewcommand{\todo}[1]{} %

\long\def\comment#1{}

\begin{document}

\title[The Lockdown Effect: Implications of the COVID-19 Pandemic on Internet Traffic]{The Lockdown Effect:\\ Implications of the COVID-19 Pandemic on Internet Traffic}

\author{Anja Feldmann}
\affiliation{
	\institution{Max Planck Institute for Informatics}
}

\author{Oliver Gasser}
\affiliation{
	\institution{Max Planck Institute for Informatics}
}

\author{Franziska Lichtblau}
\affiliation{
	\institution{Max Planck Institute for Informatics}
}

\author{Enric Pujol}
\affiliation{
	\institution{BENOCS}
}

\author{Ingmar Poese}
\affiliation{
	\institution{BENOCS}
}

\author{Christoph Dietzel}
\affiliation{
	\institution{DE-CIX}
	\institution{Max Planck Institute for Informatics}
}

\author{Daniel Wagner}
\affiliation{
	\institution{DE-CIX}
}

\author{Matthias Wichtlhuber}
\affiliation{
	\institution{DE-CIX}
}

\author{Juan Tapiador}
\affiliation{
	\institution{Universidad Carlos III de Madrid}
}

\author{Narseo Vallina-Rodriguez}
\affiliation{
	\institution{IMDEA Networks\\ICSI}
}

\author{Oliver Hohlfeld}
\affiliation{
	\institution{Brandenburg University of Technology}
}

\author{Georgios Smaragdakis}
\affiliation{
	\institution{TU Berlin\\Max Planck Institute for Informatics}
}
\renewcommand{\shortauthors}{Feldmann et al.} %

\begin{abstract}

  Due to the \covid pandemic, many governments imposed lockdowns that forced
  hundreds of millions of citizens to stay at home. The implementation of 
  confinement measures increased Internet traffic demands of
  residential users, in particular, for remote working, entertainment,
  commerce, and education, which, as a result, caused traffic shifts 
  in the Internet core.

  In this paper, using data from a diverse set of vantage points (one ISP,
  three IXPs, and one metropolitan educational network), we examine the effect of
  these lockdowns on traffic shifts.  We find that the traffic volume increased
  by 15-20\% almost within a week---while overall still modest, this
  constitutes a large increase within this short time period. However, despite this surge,
  we observe that the Internet
  infrastructure is able to handle the new volume, as most traffic shifts occur
  outside of traditional peak hours. When looking directly at the traffic sources, it turns out
  that, while hypergiants still contribute a significant fraction of traffic, we
  see \one a higher increase in traffic of non-hypergiants, and \two traffic
  increases in applications that people use when at home, such as Web conferencing, VPN, and gaming.
  While many networks see increased traffic
  demands, in particular, those providing services to residential users, 
  academic networks experience major overall decreases. Yet, in these networks,
  we can observe substantial increases when considering  
  applications associated to remote working and lecturing. 

\end{abstract}

\setlength{\TPHorizModule}{\paperwidth}
\setlength{\TPVertModule}{\paperheight}
\TPMargin{5pt}
\begin{textblock}{0.8}(0.1,0.02)
    \noindent
    \footnotesize
    If you cite this paper, please use the IMC reference:
    Anja Feldmann, Oliver Gasser, Franziska Lichtblau, Enric Pujol, Ingmar Poese, Christoph Dietzel, Daniel Wagner, Matthias Wichtlhuber, Juan Tapiador, Narseo Vallina-Rodriguez, Oliver Hohlfeld, and Georgios Smaragdakis. 2020.
    The Lockdown Effect: Implications of the COVID-19 Pandemic on Internet Traffic.
    In \textit{Internet Measurement Conference (IMC '20), October 27--29, 2020, Virtual Event, USA.}
    ACM, New York, NY, USA, 18 pages.
    https://doi.org/10.1145/3419394.3423658
\end{textblock}

\maketitle

\section{Introduction}\label{sec:intro}
The profile of a typical residential user---in terms of bandwidth usage and
traffic destinations---is one of the most critical parameters that
network operators use to drive their network operations and inform
investments~\cite{Edge-delay:2018,Maier:IMC2009,ISP-traffic-TNSM}. In the last twenty years, user profiles
have changed significantly. We observed user profile shifts from peer-to-peer
applications in the early 2000s~\cite{sigcomm2011-bittorent,Glasnost10,P4P}, to content delivery and streaming applications in
2010s~\cite{Arbor:SIGCOMM2010,Conviva-user-engagement,Akamai-Video-Stream-Quality-effect:IMC2012,CoNEXT2019-HG,Labovitz-2009-2019}, and more recently to mobile
applications~\cite{Cellular-sigmetrics2011,LTE:sigcomm2013}. Although changes in user profiles are a
moving target, they typically have time scales of years. Thus, staying up to
date, \eg via measurements, was feasible.

\begin{figure}[!t]
	\centering
	\includegraphics[width=0.45\textwidth]{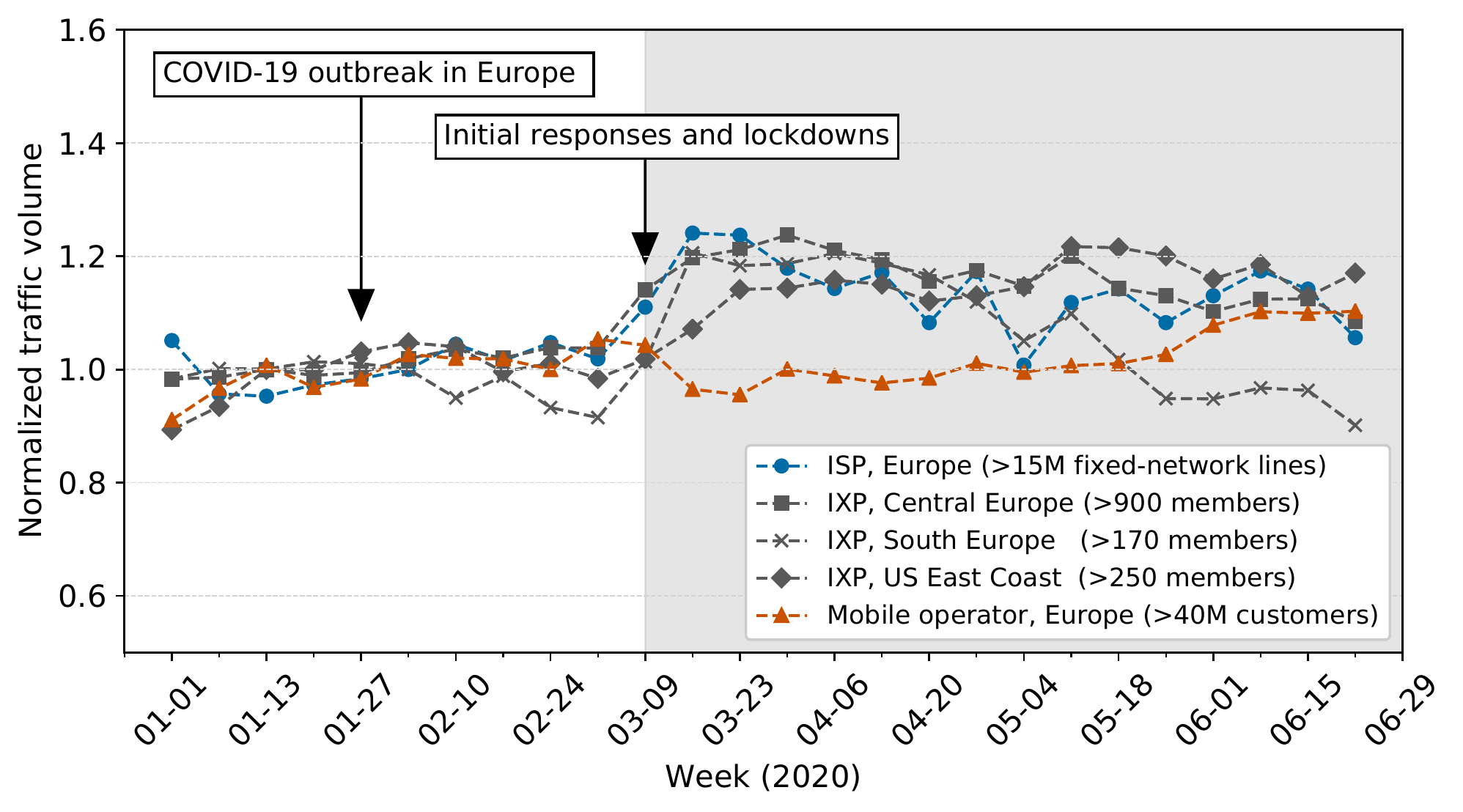}
	\caption{Traffic changes during 2020 at multiple vantage points---daily traffic averaged per week normalized by the median traffic volume of the first up to ten weeks.}
    \label{fig:overview}
  \vspace*{-1.2em}
\end{figure}

\begin{figure*}[!bt]
    \begin{subfigure}[t]{0.32\textwidth}
        \includegraphics[width=\textwidth]{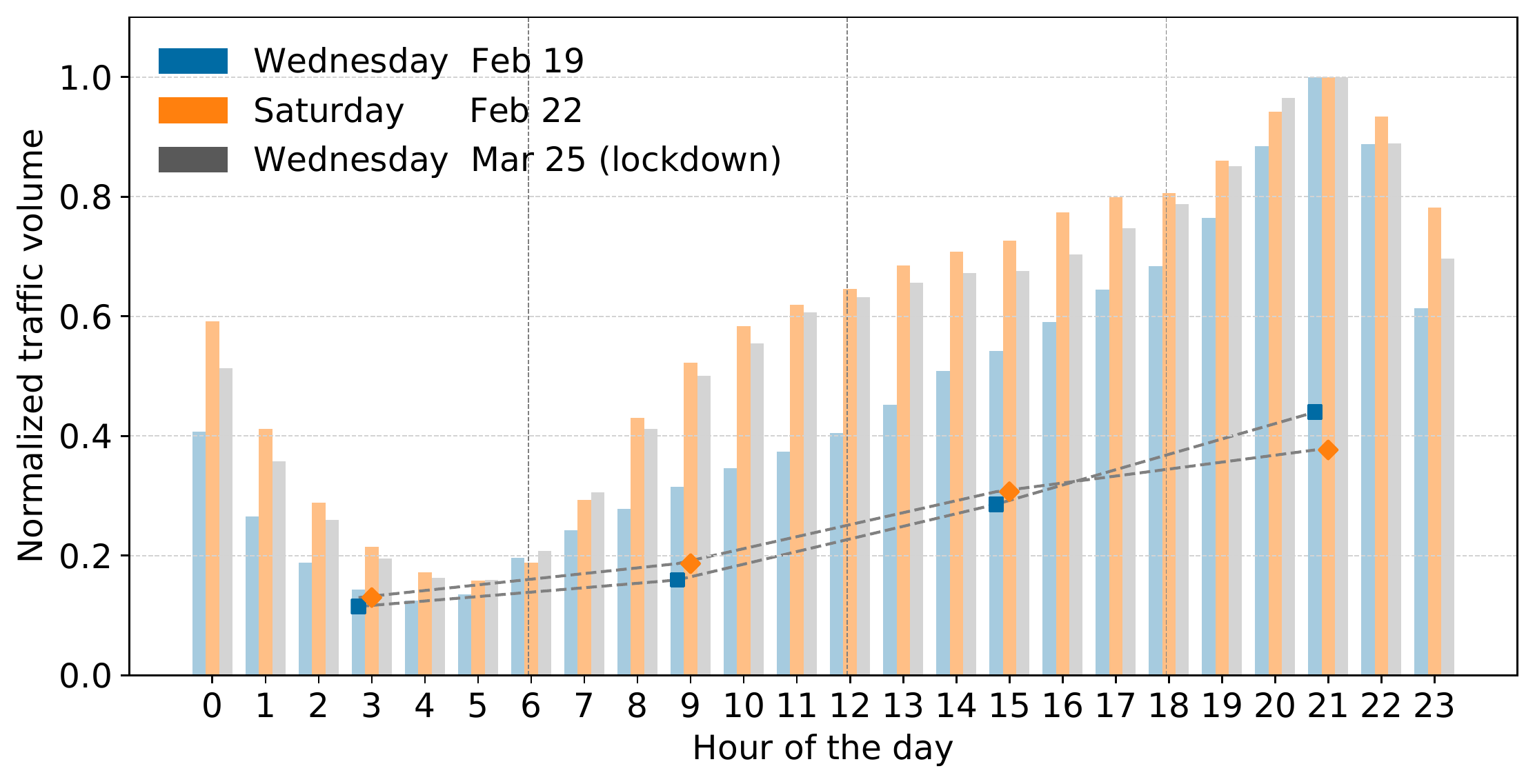}
        \caption{\ispS: Hourly traffic increase and workday vs.\ weekend pattern for February 19 (Wed), February 22 (Sat), March 25
        (Wed).}
        \label{fig:workday-vs-weekend}
    \end{subfigure}
    \hspace{0.15em}
    \begin{subfigure}[t]{0.32\textwidth}
        \includegraphics[width=\textwidth]{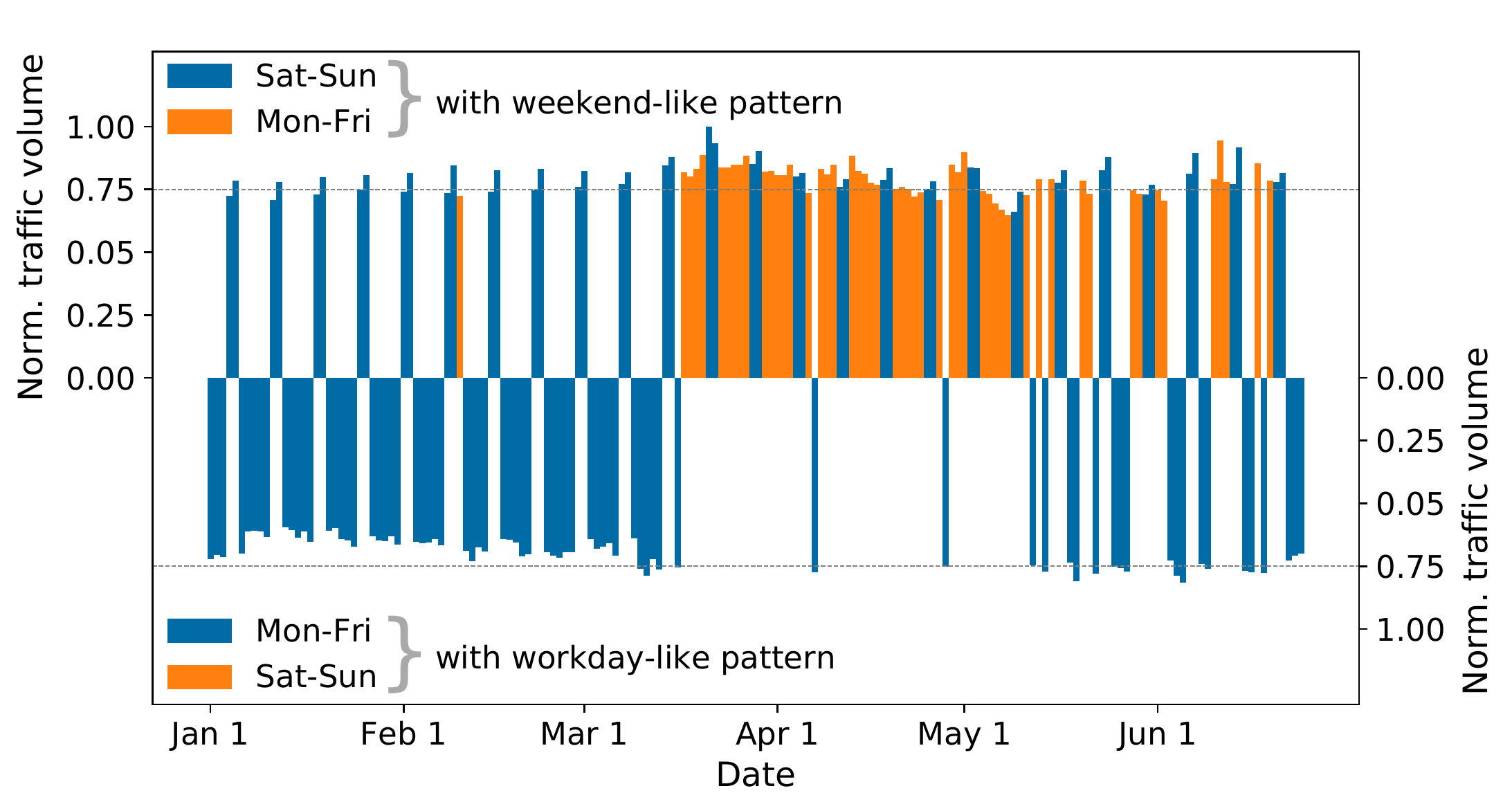}
        \caption{\ispS: Workday-like (bottom) vs. weekend-like (top) January 1--June 24.}
        \label{fig:dayClassificationISP}
    \end{subfigure}
    \hspace{0.15em}
    \begin{subfigure}[t]{0.32\textwidth}
        \includegraphics[width=\textwidth]{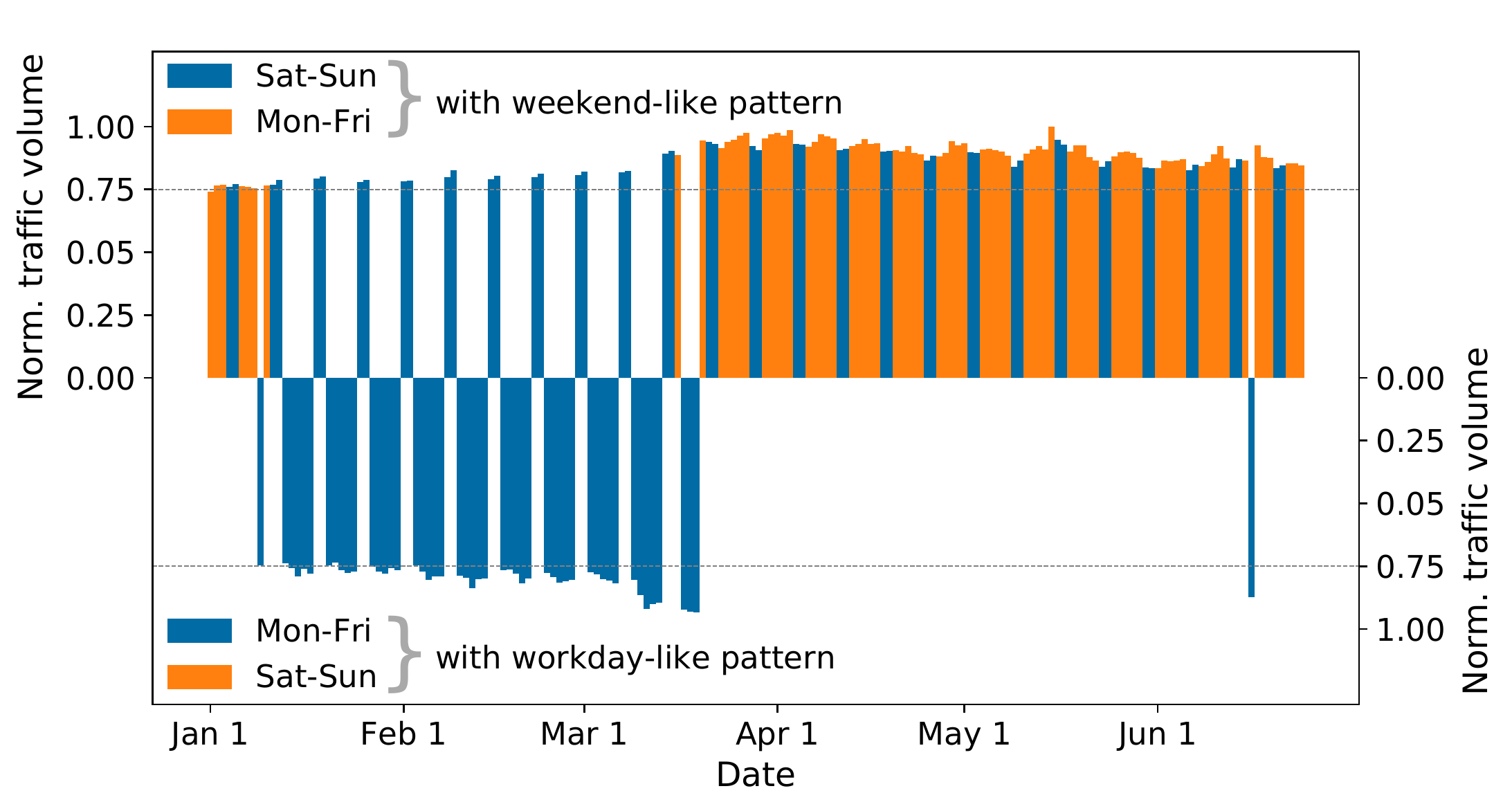}
        \caption{\ixpceS: Workday-like (bottom) vs. weekend-like (top) January 1--June 24.}
        \label{fig:dayClassificationIXP}
    \end{subfigure}
    \vspace*{-1em}
    \caption{Drastic shift in Internet usage patterns for times of day and
    weekends/workdays.}\label{fig:dayClassification}
    \vspace*{-1em}
\end{figure*}

The \covid pandemic is most likely a once in a generation global phenomenon that
drastically changed the habits of millions of Internet users around the globe.
As a result of the government mandated lockdowns, a large fraction of the population had to depend on their
residential Internet connectivity for work, education, social activities, and entertainment. Unexpectedly, the
Internet held up to this unforeseen demand~\cite{wp-internet-works}
with no reports of large scale outages or failures in more developed countries. 
This unique phenomenon allows us to observe changes that may be expected within months or years
in a matter of days.

\afblock{\covid-induced weekly growth.} We observe a significant traffic evolution in 2020 at multiple Internet vantage points in Figure~\ref{fig:overview}.
The \covid outbreak reached Europe
in late January (week 4) and first lockdowns were imposed in
mid March (starting on week 11). Thus, we normalize weekly traffic volumes by the median traffic volume of the first ten weeks of 2020 (pre-lockdown period). We can clearly identify drastic changes in the data collected at multiple and diverse vantage points (see
Section~\ref{sec:datasets} for details): Traffic demands
for broadband connectivity, as observed at an \isp %
as well as at a major \ixpce%
and an \ixpse%
increased
slowly at the beginning of the outbreak and then more rapidly by more than 20\% after the lockdowns started. The traffic
increase at the \ixpus%
trails the other data sources since the lockdown occurred several weeks later. While we observe this
phenomenon at the ISP and IXP vantage points, one difference between them is that the relative traffic increase at the
IXP seems to persist longer while traffic demand at the ISP decreases quickly towards May. This correlates with the
first partial opening of the economy, including shop reopenings in this region
in mid-April and further relaxations
including school openings in a second wave in May. Our findings are aligned with
the insights offered by mobility reports published by
Google~\cite{Google-mobility-pandemic} and the increased digital demand as
reported by Akamai~\cite{Akamai-30pcUS-pandemic,Akamai-30pcEU-pandemic},
Comcast~\cite{Comcast-covid19},
Google~\cite{Google-covid19},
Nokia Deepfield~\cite{Labovitz-covid-19},
and TeleGeography~\cite{Telegeography-covid19}.

\afblock{Drastic shift in usage patterns.} In light of the global \covid pandemic a total growth of traffic is somewhat expected. More relevant for the operations of networks is how exactly usage patterns are shifting, \eg, during the day or on different days of a week. To this end, we show the daily traffic patterns at two of the above mentioned vantage points in Figure~\ref{fig:dayClassification}. The Internet's regular workday traffic patterns are significantly different from weekend
patterns~\cite{structural-analysis-WDWE,BLINC,mobile-WDWE}. On workdays, traffic
peaks are concentrated in the evenings, see Figure~\ref{fig:workday-vs-weekend}. For instance, Wed., February 19 vs. Sat., February 22, 2020: With the
pandemic lockdown in March, this workday traffic pattern shifts towards a
continuous weekend-like pattern, as can be seen in 
the daily pattern for Mar. 25, 2020 in Figure~\ref{fig:workday-vs-weekend}. More specifically, we call a traffic pattern a workday pattern if the traffic spikes in the evening hours and a weekend pattern if its main activity
gains significant momentum from approximately 9:00 to 10:00 am. For our classification, we
use labeled data from late 2019 and use an aggregation level of 6 hours. Then, we apply this
classification to all available days in 2020. Figures~\ref{fig:dayClassificationISP}
and~\ref{fig:dayClassificationIXP} show the normalized traffic for days
classified as weekend-like on the top and for workday-like
on the bottom. If the classification is in line with the actual day (workday or weekend) the bars are colored blue, otherwise they are colored in orange.
We find that up to mid-March, most weekend days are
classified as weekend-like days and most workdays as workday-like days. The only exception
is the holiday period at the beginning of the year in
Figure~\ref{fig:dayClassificationIXP}. This pattern changes drastically 
once the confinement measures are
implemented: Almost all days are classified as weekend-like. This change persists in Figure~\ref{fig:dayClassificationIXP} until the end of June due to the vacation period, which is consistent with the behavior observed in 2019 (not shown). In contrast, Figure~\ref{fig:dayClassificationISP} shows that the shift towards a weekend-like pattern becomes less
dominant as countermeasures were relaxed in mid-May.

These observations raise the question of the cause for this significant
traffic growth and shift in patterns, given that many people are staying at
home for all purposes, \eg working from home,
remote education, performing online social activities, or consuming entertainment content. The increased
\emph{demand in entertainment},
\eg video streaming or gaming,
may imply an increase in hypergiant traffic. This is in accordance with a
statement by a commissioner of the European Union which stated that major streaming companies
reduced their video resolution to the standard definition from March 19, 2020
onward~\cite{reported-date-for-netflix,eu-streaming-reduction}. According to mainstream media, some started to upgrade their services back to high definition or 4K around May 12, 2020~\cite{netflix-restriction-lift}.
Furthermore, \emph{the need for remote working} may imply an increased demand for VPN services, usage of video conference systems, email, and
cloud services.

\begin{figure*}[tb]
	\centering
	\begin{subfigure}[t]{0.49\textwidth}
		\centering
		\includegraphics[width=0.9\textwidth]{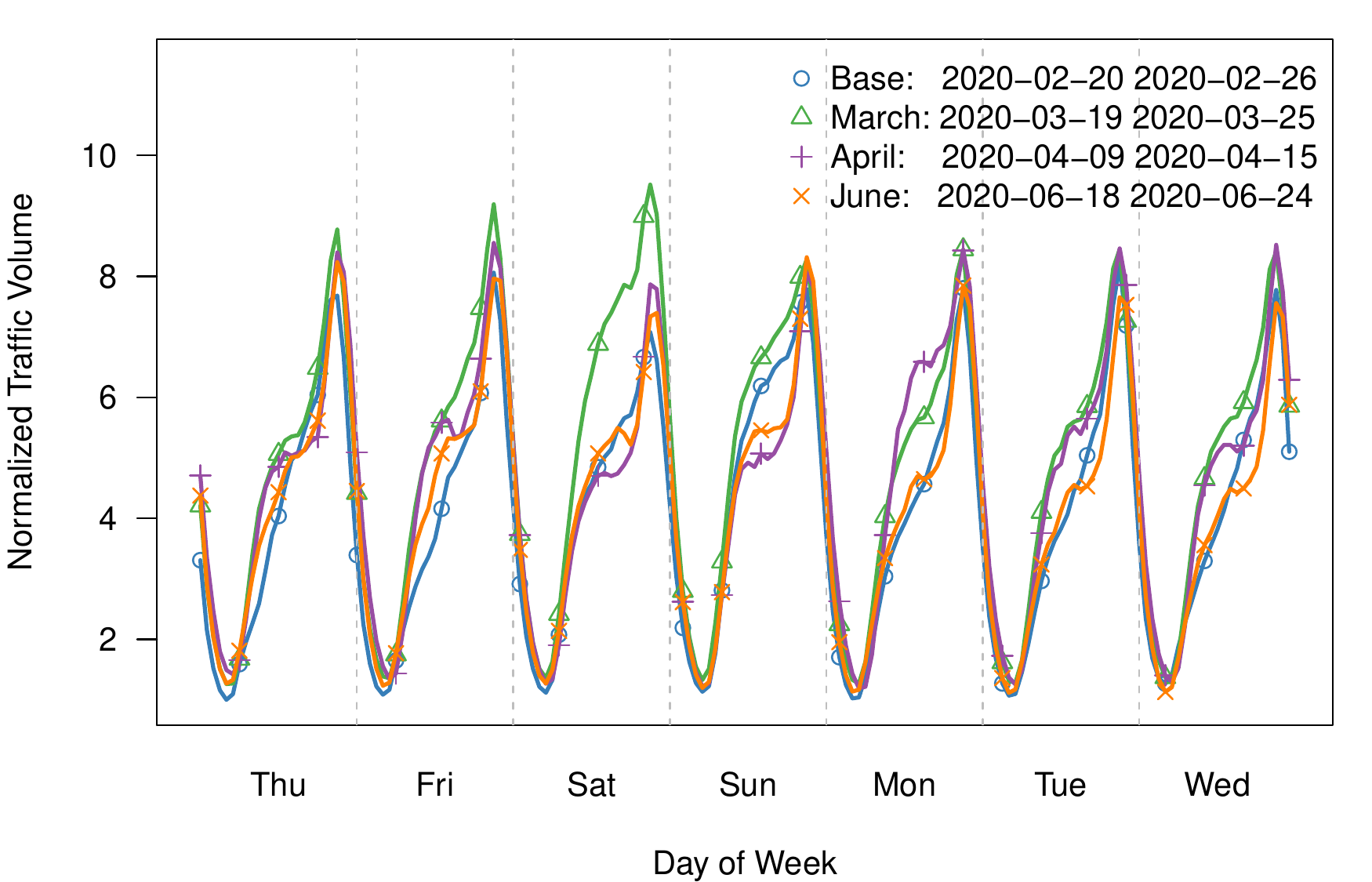}
		\caption{L-ISP (Central Europe).}
		\label{fig:aggregated-traffic-timeseries-isp}
		\label{fig:dayPattern-isp}
	\end{subfigure}%
	\begin{subfigure}[t]{0.49\textwidth}
		\centering
		\includegraphics[width=0.9\textwidth]{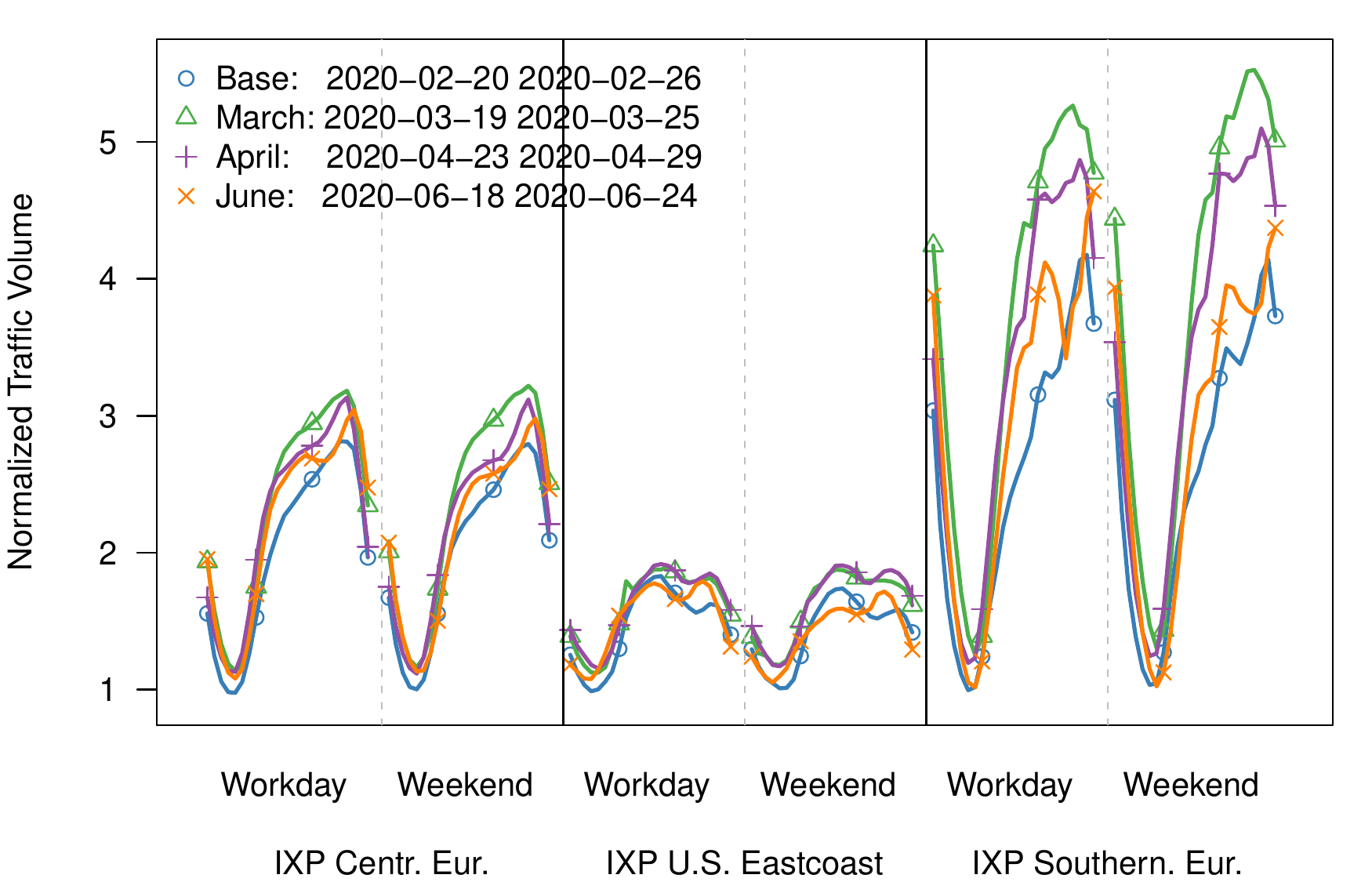}
		\caption{IXPs (Central Europe/US Eastcoast/Southern Europe).}
		\label{fig:aggregated-traffic-timeseries-ixp}
		\label{fig:dayPattern-ixp}
	\end{subfigure}
        \vspace*{-1em}
	\caption{Time series of normalized aggregated traffic volume per hour
          for \ispS and three IXPs for four selected
          weeks: before, just after, after, and well after
          lockdown (base/March/April/June).}
	\label{fig:aggregated-traffic-timeseries}
\end{figure*}
In this paper, we study the effect that government-mandated lockdowns had on the
Internet by analyzing network data from a major
Central European ISP (\ispS), three IXPs located in 
Central Europe, %
Southern Europe, %
and the US East Coast, %
and an Spanish educational network (\uniS). %
This enables us to \emph{holistically} study the effects of the \covid
pandemic both from the network edge (\ispS/\uniS) and the Internet core
(IXPs). We find that:

\begin{itemize}[leftmargin=*]
\item Relative traffic volume changes follow user changing
  habits---causing ``moderate'' increases of 15-20\% at the height of the
  lockdown for the ISP/IXPs, but decreases up to 55\% at the \uniS network. Even
  after the lockdown is scaled back, some of these trends remain: 20\%
  at the \ixpceS but only 6\% at the \ispS.
\item Most traffic increases happen during non-traditional peak hours.
  Daily traffic patterns are moving to weekend-like patterns.
\item Online entertainment demands account for hypergiant traffic surge.
Yet, the need
for remote working increases the relative traffic share of many ``essential''
  applications like VPN and conferencing tools by more than
  200\%.  At the same time, the traffic share for other traffic classes decreases
  substantially, \eg traffic related to education, social media, and---for some periods---CDNs.
\item At the IXP-level, we observe that port utilization increases. This
phenomenon is mostly explained by a higher traffic demand from residential 
users.
\item Traffic changes are diverse, increasing in some network ports while
decreasing in others. 
  One example of the latter is the \uniS network, where
  we observe a significant drop in traffic volume on workdays after the
lockdown measures loosened,
  with a maximum decrease of up to 55\%. Yet, remote working and lecturing
  cause a surge in incoming traffic, \eg for email and VPN
  connections. The \uniS traffic shift is antagonistic,
  yet complementary, to the observations made in other vantage points.
\end{itemize}

\comment{
We expect an increase in VPN of video conferencing,

Home office traffic vs. the rest of the traffic
  - home office traffic is actually not that much....
  - yet the traffic is more diverse in terms of destinations....
  - involves more upstream than downstream

Local connectivity sufficient? We cannot answer...
  - traffic volume per IP increase:  before vs. after plotted as two CDFs
     -- Look at the distributions in particular upstream

Connectivity from the Hypergiants sufficient?
  - traffic volume per Hypergiant and their increase
  - impact of resolution change....

Need for increase in capacity?
  - IXP sees XXX increase in port capacity
  - Enterprise connectivity?

if these changes in traffic patterns are caused by the
likely increased need for entertainment and information or by the increased
need to support home office.

}

\comment{
In the process, we
consider the new mix of applications that contribute to the new user profile
being residential, enterprise, and mobile. We also characterize the
increasingly popular applications and how they are delivered to the end-users;
many of these applications are cloud-based and non-elastic and, thus,
non-cacheable. We also investigate to what extend the flattening of the
Internet topology and content provider's investments, observed in the last
years, contribute to the smooth operation of the access and core network
despite the rapid user profile changes.
}

\section{Datasets}\label{sec:datasets}

This section describes the network traffic datasets that we used for our analysis.
We utilize vantage points at the core of the Internet (IXPs), at the backbone and peering points of a major
Internet Service Provider, and at the edge (a metropolitan university network), all which we will describe below.

\afblock{\ispS:} Network flows from a large Central European ISP that provides service to more than 15 million fixed line subscribers and also operates a transit network (Tier-1). The ISP does not host
content delivery servers inside its network, but it has established a large
number of peering agreements with all major content delivery and cloud
networks at multiple locations.  This ISP uses \nf~\cite{Cisco-Netflow} at
all border routers to support its internal operations.
We rely on two different sets of NetFlow records for this paper. First, we
use NetFlow data collected at ISP's Border Network Gateways~\cite{CISCO-BNG} to
understand the impact of changing demands of the ISPs' subscribers.
Second, we use NetFlow records collected at the ISP's border routers to gain a
better understanding about how companies running their own ASNs are
affected by these changes.

\afblock{IXPs:} Network flows from the public peering platform of three major
Internet Exchange Points (IXPs). The first one has more than 900 members, is
located in Central Europe (\ixpceS) and has peak traffic of more than 8
Tbps. The \ixpceS is located in the same country as the \ispS. The second one has
more than 170 members, is located in Southern Europe (\ixpseS) and has a peak
traffic of roughly 500 Gbps. It covers the region of the \uniS network. The third one has 250 members, is located at the
US East Coast (\ixpusS) and has a peak traffic of more than 600 Gbps. At the IXPs
we use IPFIX data~\cite{rfc7011}.

\afblock{\uniS:} Network flows from the REDImadrid~\cite{REDImadrid} %
academic network interconnecting 16 independent
universities and research centers in the region of
Madrid.  It serves nearly 290,000 users
including students, faculty, researchers, student halls, WiFi networks
(including Eduroam),
and administrative and support staff. The network operator provided us with anonymized \nf data
captured at their border routers (captured at all ingress interfaces) during 72
days in the period of Feb 28 to May 8, 2020.
The final dataset contains
5.2B flows entering or leaving the educational network.

We augment our analysis with \nf records from a large mobile
operator that operates in Central Europe, with more than 40 million customers.

\comment
{ %
\afblock{Roaming Network:} \nf records from an Internetwork Packet Exchange
located in the same metropolitan area as \ispS that mobile operators utilize to
exchange traffic for mobile data roaming.
}

\afblock{Normalization:} Since all data sources exhibit vastly differing
traffic characteristics and volumes, we normalize the data to make it
easier to compare.  For plots where we show selected weeks only, we
normalize the traffic by the minimum traffic volume.  For plots spanning a
larger timeframe, we normalize the traffic by the median traffic volume of
the first ten weeks of 2020, depending on the availability of data.

\afblock{Time frame:} We use two methods to reflect the developments since the beginning of the COVID pandemic: (a) for
general trends over time we use continuous data from \emph{Jan 1, 2020---Jun 24, 2020}, (b) to highlight detailed
developments we compare 7-day periods as shown in Table~\ref{tab:dates} from before, during, after and well after the
lockdown in 2020.\footnote{Due to data availability, the \ispS is using Apr 09--15 which
    covers the Easter holiday period. As partial lockdowns and travel restrictions were still in place, the introduced bias may be very small.}

\begin{table}
\footnotesize
\begin{tabular}{l | c c c c c}
    \toprule
                 &\textbf{\ispS} & \textbf{\ixpceS} & \textbf{\ixpseS} & \textbf{\ixpusS} & \textbf{EDU} \\
    \midrule
    \emph{base}  & Feb 20--26 &Feb 20--26 &Feb 20--26&Feb 20--26&Feb 20--26\\
    \emph{March} &Mar 19--25&Mar 19--25&Mar 12--18&Mar 19--25&Mar 12--18\\
    \emph{April} & Apr 09--15&Apr 23--29&Apr 23--29&Apr 23--29&Apr 23--29\\
    \emph{June}  &Jun 18--24&Jun 18--24&Jun 18--24&Jun 18--24& n/a \\
    \bottomrule
\end{tabular}
\caption{Summary of the dates used in weekly analyses. Dates in Southern Europe vary due to different courses of the
pandemic.}
\label{tab:dates}
\end{table}

\begin{figure*}[tb]
    \centering
    \begin{subfigure}[b]{0.49\textwidth}
        \centering
        \includegraphics[width=0.99\textwidth]{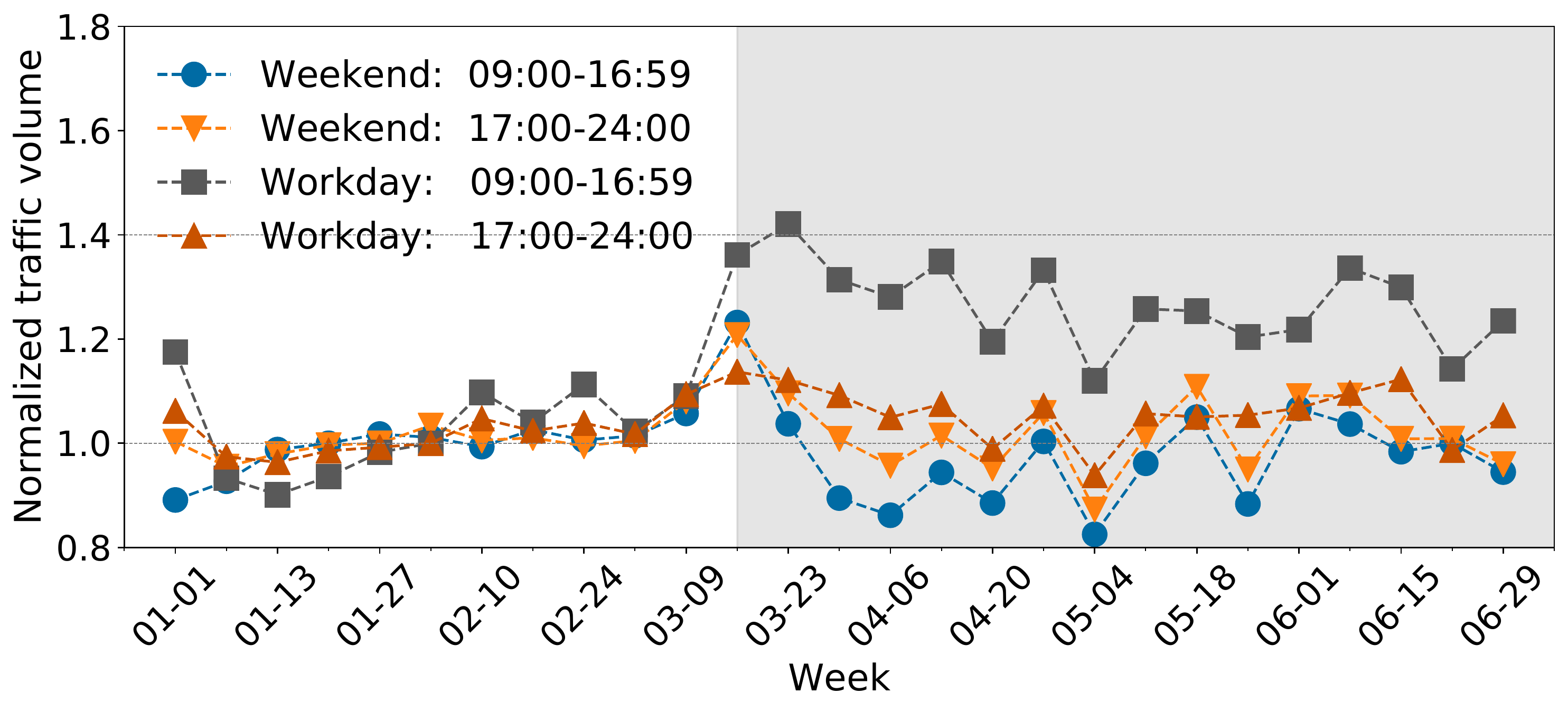}
        \caption{Hypergiants}
        \label{fig:hypergiant-other-weekday-workhours}
    \end{subfigure}
    \begin{subfigure}[b]{0.49\textwidth}
        \centering
        \includegraphics[width=.99\textwidth]{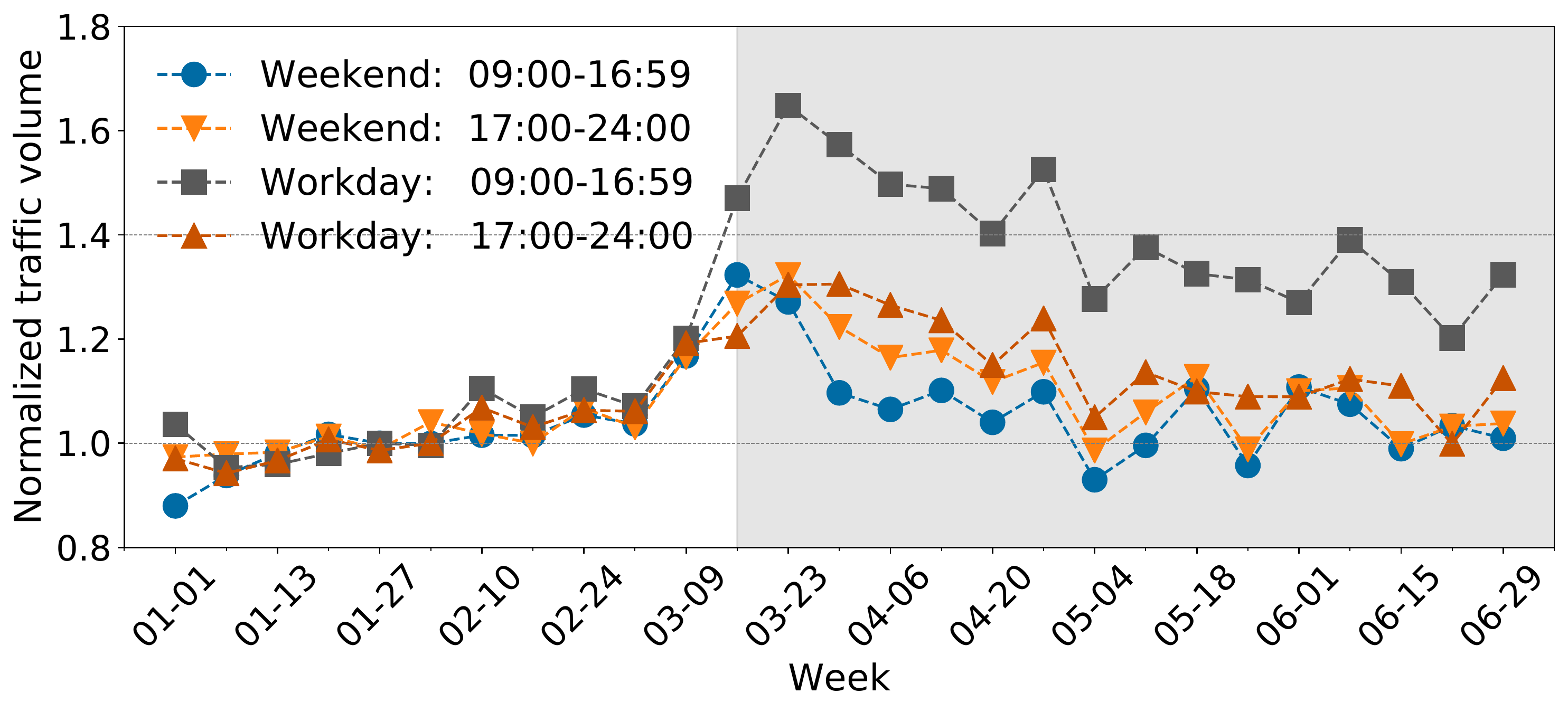}
        \caption{Other ASes}%
        \label{fig:hypergiant-other-weekday-offhours}
    \end{subfigure}
    \vspace*{-1em}
    \caption{\ispS: Normalized daily traffic growth for hypergiants vs.\ other ASes
        across time.}
    \label{fig:hypergiant-other}
\end{figure*}

\comment{
\subsection{Metadata}

\afblock{Prefix characterization:} Network operators tagged the announced
BGP prefixes of their networks as residential or enterprise, making it easier for us to analyze on residential 
user behavior during the lockdown. In addition, we characterize prefixes of other networks based on the
business type and online products each network (AS) offers.  We tag prefixes as
Video on Demand (VoD), (Cloud-) Gaming, Messaging and Social Media,
Collaborative Working, CDN, and Research/Educational Networks.
\smallskip

\afblock{Peering data:} The ISP operators maintain a detailed list of the
peering agreements including associated capacities and peering locations. The
IXP operator also maintains a list of port assignment to members and the
capacity of each port. Bilateral and multilateral peering can be inferred from
the BGP messages exchanged between members as well as from the IXP routing
server.  Using the above information it is possible to infer the level of
utilization of peering and ports at all our vantage points. It is also possible
to report on the growth of peering and capacity during the period of our
study.
\smallskip

\afblock{Lock down dates:} We also collected information related to the pandemic, \eg lock down dates, number of COVID-19 cases, at all the
locations the flow data is collected at. %
}

\subsection{Ethical Considerations}

Both \nf and IPFIX data provide only flow summaries based on the packet
header and do not reveal any payload information. To preserve users privacy, all data
analyses are done on servers located at the premises of the ISP, IXPs, and the
academic network. IP addresses are hashed to prevent information leaks and
raw data being transferred. The output of the analyses are the aggregated
statistics as presented in the paper. The data at the ISP and IXPs is collected
as a part of their routine network analysis. For obtaining and analyzing the
academic network data (EDU), we obtained IRB approval from the respective
institutions.
\section{Aggregated Traffic Shifts}\label{sec:aggregated-traffic}

To understand traffic changes during the lockdown we first look for overall traffic
shifts before, during, and after the strictest lockdown periods. Moreover, we take
a look at hypergiant ASes vs.\ other ASes, shifts in link utilization, and ASes relevant for remote working.

\subsection{Macroscopic Analysis}
\label{sec:macroperspective}

Figure~\ref{fig:aggregated-traffic-timeseries} plots the aggregated normalized 
traffic volume in bytes at the granularity of one hour for the \ispS, \ixpceS, 
\ixpusS, and \ixpseS in four selected weeks (see Table~\ref{tab:dates}).
For the \ispS,
Figure~\ref{fig:aggregated-traffic-timeseries-isp} shows the time series using
normalized one-hour bins. For the IXPs,
Figure~\ref{fig:aggregated-traffic-timeseries-ixp} reports the hourly average
for workdays and weekends. 

First of all, we see that the overall traffic after the lockdown increased by
more than 20\% for the \ispS and 30\%/12\%/2\% for the
\ixpseS/\ixpceS/\ixpusS, respectively. Once the lockdown measures were relaxed,
the growth started declining for the \ispS but persisted for the \ixpceS and the \ixpseS.
These differences are most likely attributed to the fact that the
\ispS traffic pattern is dominated by end-user and small enterprise
traffic---recall, we are not analyzing any transit traffic---while the \ixpceS
has a wider customer base. Traffic persistently increased for the \ixpusS
where the lockdown was put into place later.

As previously noted, the \ispS time series shows the same workday to weekend
traffic patterns shifts starting with the lockdown in mid-March. In accordance with
that observation, traffic increases much earlier in the day with a small dip at
lunchtime. However after lunch hours, traffic grows to roughly the same volume during the evening
time, spiking late in the evening. This change persists throughout the
lockdown. Once this was relaxed, the pattern became less pronounced and the shift
to a weekend like pattern became less dominant.
Additionally, it is important to note 1) the Easter vacations in the April week, and 2) the seasonal effects in the weekend of the June week (an increase of outdoor activities). 

For all IXPs, see Figure~\ref{fig:aggregated-traffic-timeseries-ixp}, not only do
we see an increase in peak traffic but also in the minimum traffic
levels. This correlates with link capacity upgrades of many IXP members
leading to overall increases of 3\% at \ixpceS, 12\% at \ixpse and 20\% at \ixpus.
In addition, we see the increase in traffic during daytime, which is
very pronounced at the \ixpceS. However, the differences between weekends and
workdays are not as apparent as at the ISP. Interestingly, as lockdown
measures were mandated, the daytime traffic again decreases but stays well
above the pre-lockdown level. In contrast, traffic at the \ixpusS barely changes
in March and increases only in April, otherwise showing similar
effects as the other IXPs. The delayed increase in volume is likely due to the later
lockdown in the US.  Overall, the effects of the time of day at this IXP are less
pronounced compared to the two others because it (a) serves customers from many
different time zones, and (b) members are diverse and
include eyeball as well as content/service providers. In contrast, the
\ixpseS interconnects more regional networks, and as such the traffic patterns
are closer to the ones of the \ixpceS.

\subsection{Hypergiants}

To understand the composition of residential traffic, we investigate who is
responsible for the traffic increase at the \ispS. The first step
is to look at the top 15
hypergiants~\cite{Arbor:SIGCOMM2010,bottger2018looking,bottger2017hypergiant}
(full list in Appendix \ref{appendix:hypergiants}).  Hypergiants are networks
with high outbound traffic ratios that deliver content to approximately millions of users
in the locations at which we have vantage points. The 15 hypergiants we consider in this study
are responsible for about 75\% of the traffic delivered to the end-users of
the \isp\, which
is consistent with recent reports in the
literature~\cite{Labovitz-2009-2019,CoNEXT2019-HG,Edge-delay:2018}. We note
that the fraction
of hypergiant traffic vs.\ traffic from other ASes does \emph{not} change drastically for
the \ispS as well as all IXPs.

Given that the overall traffic has increased, we next report the
relative increase of the two AS groups compared to the median traffic volume
during the pre-lockdown period, see
Figure~\ref{fig:hypergiant-other}. In detail, we focus on different times of
day and days within the week. We find that the relative traffic increase is
\emph{significantly} larger for \emph{other ASes} than for hypergiants.

Both sets of time series are more or less on top of each other until the lockdown.
This observation also holds for data from 2019 (not shown). However, after
the lockdown, the time series for the other ASes present higher deviations from the reference
value than those of the hypergiants. The most visually striking difference occurs during
working hours of work-days: Hypergiants experience a 40\% increase whereas the remaining
ASes grow by more than 60\%. While this difference is significantly reduced around mid-May,
the relative increase for both sets of ASes is still substantial.
In fact, except for the working hours during work-days, the traffic surge seems to
normalize around mid-May, especially for \emph{other ASes}. Notice the fluctuations during
weekends mornings starting around the end of April---they can be also observed in 2019 (not shown).

A plausible explanation for the increase of daily traffic volumes in this 
vantage point are family members being forced to continue 
their professional and educational activities from home. 
Yet, the demand for entertainment content---mainly video streaming---explains 
the increase in traffic volume associated with hypergiants, many of which
offer such services. The increase in
traffic by the other ASes has more facets and it requires a more thorough
analysis that incorporates traffic classification methods. 
Before doing that, 
the next subsections investigate the impact that these ASes have on parts of 
the infrastructure of some of our vantage points.

\subsection{Link Utilization Shifts}
\label{subsec:link-utilization}
\begin{figure}[tb]
	\centering
	\begin{subfigure}[b]{0.4\textwidth}
		\centering
		\includegraphics[width=\textwidth]{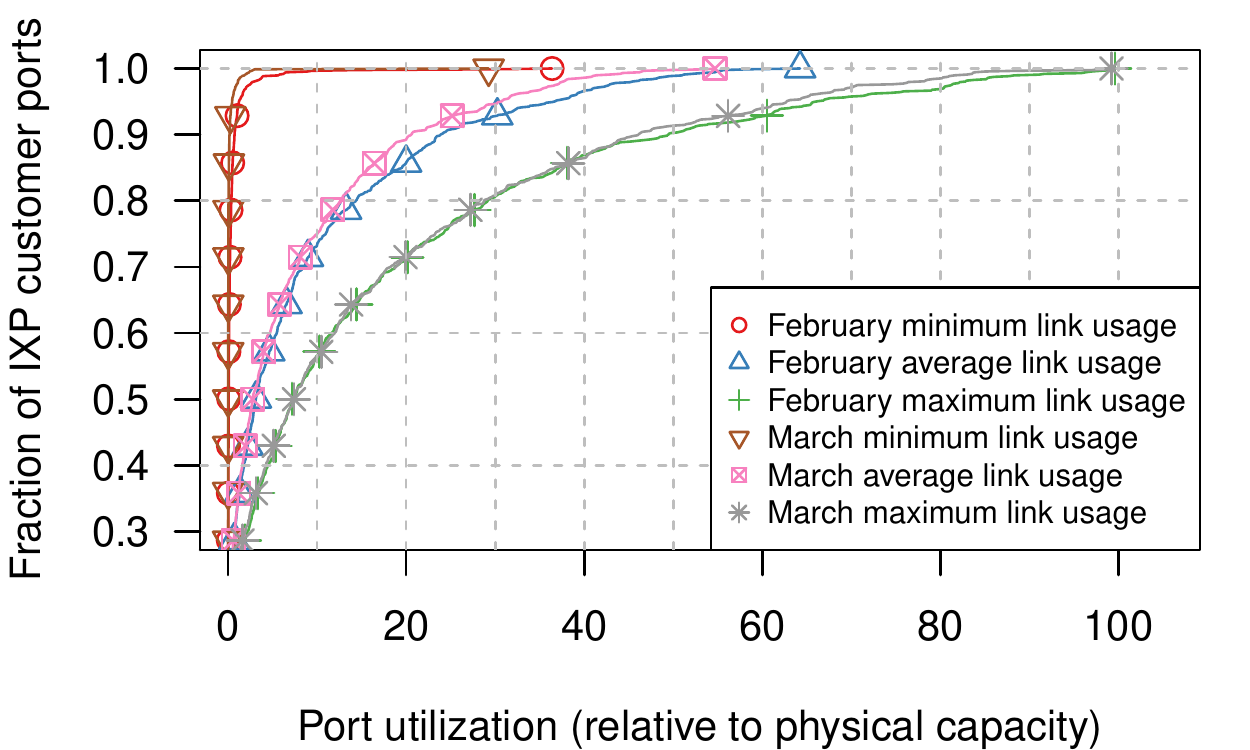}
	\end{subfigure}
    \caption{\ixpceS: ECDF of link utilization before and during the lockdown.}
	\label{fig:shift-link-utilization-ixp}
\end{figure}

We analyze to which extent the observed changes are reflected in our link utilization
dataset to assess how many networks suffer changes in their traffic
characteristics.
For this, we look at changes in relative link utilization between the
base week in February and the selected week in March. We choose \ixpceS as
reference vantage point as it houses the greatest variety of connected ASes,
thus allowing a more complete and meaningful analysis.
Our dataset reflects link capacity upgrades as well as customers switching to PNIs.
We plot the minimum, average and maximum link utilization for all members at \ixpceS in
Figure~\ref{fig:shift-link-utilization-ixp}.  
Appendix \ref{appendix:link-utilization} provides additional figures comparing 
link utilization in other months.

Figure~\ref{fig:shift-link-utilization-ixp} shows a slight shift to the left 
during lockdown. This denotes a tendency towards
decreased link usage across many IXP members which could be caused by 
link capacity upgrades or members switching to PNIs in response to increased 
traffic demand~\cite{Labovitz-covid-19}.
It is important to note that increased link usage of a network can be concealed 
by another network upgrading its port. However, the main takeaway is that many of
the non-hypergiant ASes show changes in their link usage due to the lockdown-induced shifts in
Internet usage. To gain a better understanding of this phenomenon, 
we reconsider the non-hypergiant ASes and
their role in the Internet for further analysis.

\subsection{Remote-work Relevant ASes}

\begin{figure}[tb]
	\centering
	\begin{subfigure}[b]{0.4\textwidth}
		\centering
		\includegraphics[width=0.99\textwidth]{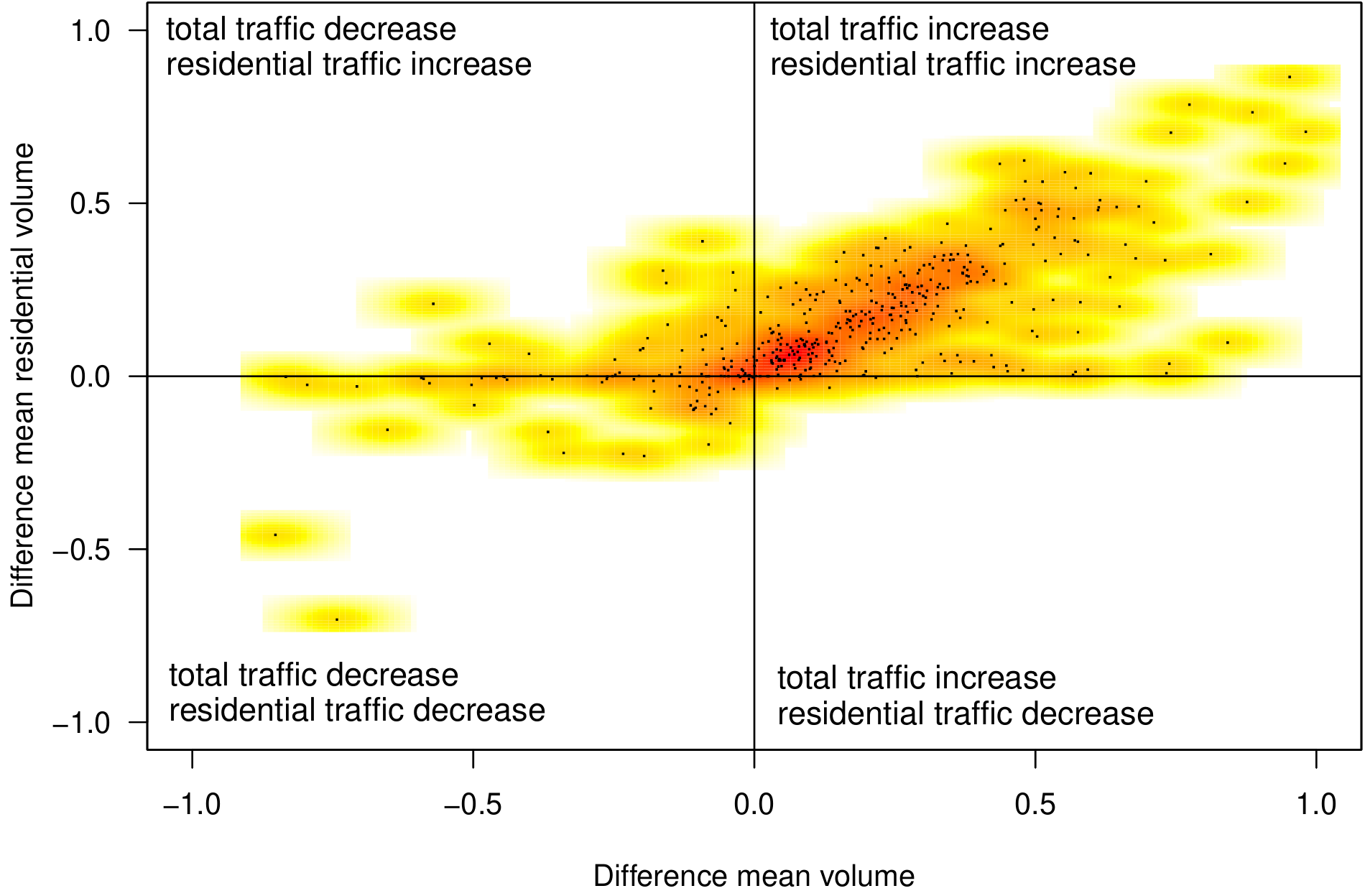}
	\end{subfigure}
	\caption{\ispS: Heatmap of traffic shift vs.\ residential traffic shift (Feb.\ vs.\ Mar.).}
	\label{fig:company-asns-s1}
\end{figure}

Having observed that the relative increase in traffic during working hours is
more pronounced for non-hypergiants ASes, we
study temporal patterns to identify which ASes are relevant
for remote work, \eg large companies with their own AS or ASes offering
cloud-based products to be used by their employees. 
To this end, we use the \isp dataset,
including its transit traffic, to compute the received and
transmitted traffic per ASN.\footnote{We are aware of limitations of this
  vantage point, \eg companies may have additional upstream providers.} 
In addition, we compute the traffic that each one of them sends and receives
to/from manually selected eyeball ASes, \ie the large broadband
providers in the region. Using this data, we define three distinct groups of
ASes: those whose traffic ratio of workday/weekend traffic is dominated by
workdays, those who are balanced, and those in which weekend traffic patterns 
dominate.

We focus on the first group, as we expect companies and enterprise subscribers 
deploying remote working solutions for their employees to fall into
this class. We crosscheck their AS numbers with the WHOIS database.
We find that a small number of content-heavy ASes also fall in this category.
In Figure~\ref{fig:company-asns-s1} we show the
difference in normalized traffic volumes between a base week in February
and one in March after
the lockdown began (x-axis) vs.\ the normalized difference in traffic from/to
eyeball ASes. We observe that some ASes experience major traffic shifts, but with
almost no residential traffic (scattered along the x-axis, and close to 0 in
the y-axis). However, for a majority of the ASes, there is a correlation
between the increase in traffic involving eyeball networks and the total
increase. This suggests that most of the traffic change is due to eyeball
networks.  Interestingly, some ASes suffer a decrease in total traffic, yet 
residential traffic grows (top-left quadrant). These are likely companies
that either offer online services that became less popular and relevant 
during the lockdown or that
do not generate traffic to the Internet ``internally''. When looking at
the other AS groups (not shown), the correlation still exists but is weaker.

These observations help us to put the implications of the lockdown measures in
perspective: Some ASes need to provision a significant amount of extra
capacity to support new traffic demands in an unforeseen fashion. 
In the following sections, we will explore which specific traffic categories 
have experienced most dramatic changes. 
\section{Transport-Layer Analysis}\label{sec:user-application-port}

Based on the overall traffic pattern shifts identified in
Section~\ref{sec:aggregated-traffic}, in this section we focus on differences
in raw transport port-protocol distributions.

\begin{figure*}[tb]
    \centering
    \includegraphics[width=\textwidth]{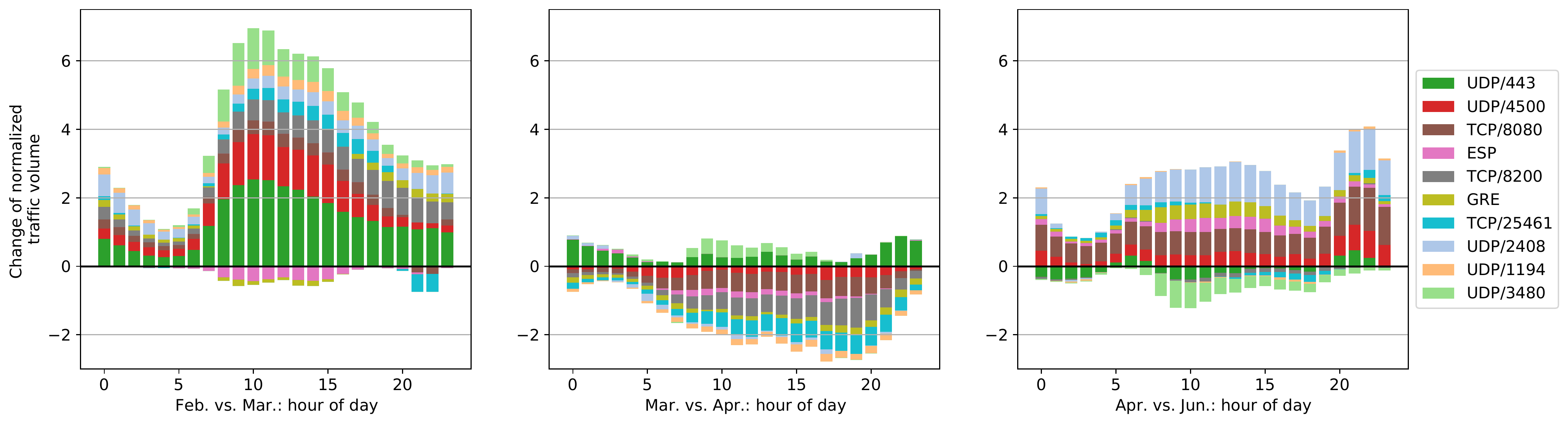}
    \caption{\ixpceS traffic difference by top application ports: normalized aggregated 
        traffic volume difference per hour comparing the workdays of February, March, April, and June.
        We omit TCP/80 and TCP/443 traffic for readability purposes.}
    \label{fig:ports-ixp}
\end{figure*}

\begin{figure*}[tb]
    \centering
    \includegraphics[width=\textwidth]{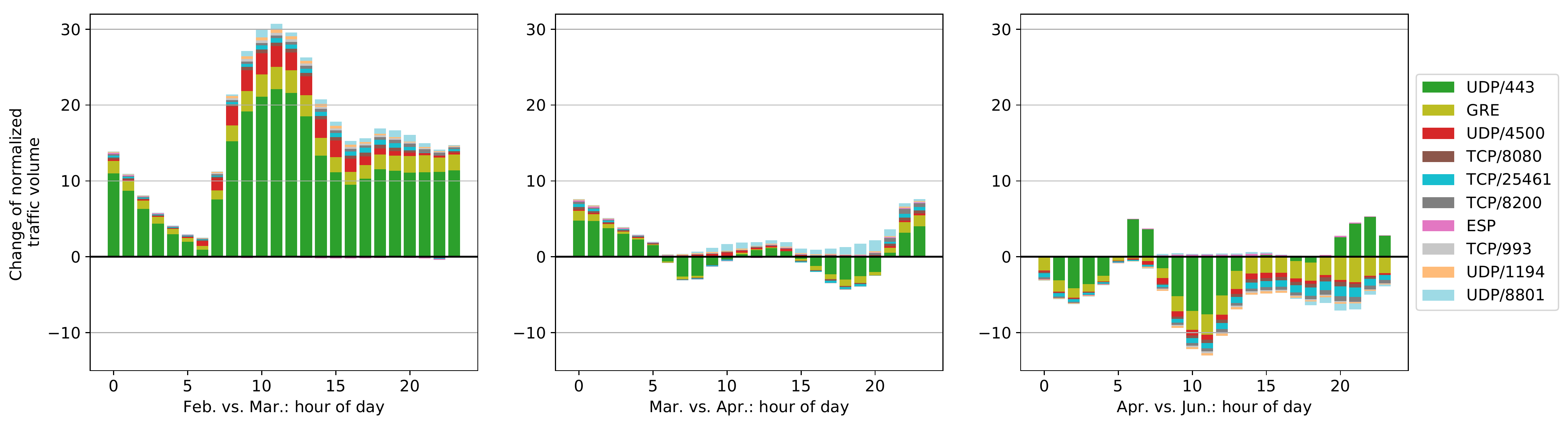}
    \caption{\ispS traffic difference by top application ports: normalized aggregated 
        traffic volume difference per hour comparing the workdays in February, March, April, and June.
        We omit TCP/80 and TCP/443 traffic for readability purposes.}
    \label{fig:ports-isp}
\end{figure*}

We analyze the shift in application traffic due to the lockdown at two vantage points, the \isp (\ispS) and the \ixpce 
(\ixpceS).
At both networks, we aggregate traffic volume statistics from four weeks described in Table~\ref{tab:dates}.
For each hour of the day, we keep separate traffic volume statistics and then compare these to the respective day and hour of the previous month, which allows us to identify diurnal patterns, and more importantly, changes therein.

We plot the top transport ports for each vantage point.
As the two most common ports TCP/443 and TCP/80 make up 80\% and 60\% of traffic at the \ispS and
\ixpceS, respectively, any small changes in their traffic volume would dominate
the plot. Therefore, we omit those from
Figures~\ref{fig:ports-ixp} and \ref{fig:ports-isp} for clarity purposes. 
\footnote{We also consider alternative HTTP
port TCP/8080, rendered in the figures, but we do not observe any 
significant change in its usage.}
We instead focus on the top 3--12 ports. 
Figure~\ref{fig:ports-ixp} depicts changes in traffic volume per transport-layer port for the \ixpceS, and Figure~\ref{fig:ports-isp} for the \ispS.
Note that we aggregate the hours of day of all working days of a week into a single subplot.
Plots for aggregated weekend days along with plots directly comparing changes to
the base week of February are shown in Appendix~\ref{appendix:application-port}.

While both networks share similar top ports, their distribution, and the changes
in these distributions over time, are very different.
This reflects the different types of customers present at these vantage points.
Recall, that the \ispS dataset consists of subscriber traffic, which is largely composed of end-users and small 
enterprises, while the \ixpceS one has a very diverse set of members across the entire Internet economy exchanging traffic over its platform.
In general, we see a very strong increase at the \ixpceS as well as at the \ispS when comparing the changes in March (leftmost subplots), compared to the more gradual changes in the following months (middle and rightmost subplots).

Next, we analyze in-depth specific ports to more accurately attribute overall changes in diurnal patterns:

\begin{description}[leftmargin=0pt]
 \item[QUIC:] Running on port UDP/443, QUIC is mainly used for streaming purposes by \eg Google and Akamai \cite{ruth2018first}.
        QUIC traffic increases 30\%--80\% at the \ispS and 
        about 50\% at the \ixpceS when comparing traffic volumes in March with the base week of February. Once the lockdown starts, we see the largest increase at the \ispS in the morning hours.
        Moreover, at the \ixpceS the increase is more gradually distributed over the day.
        This likely reflects the behavior of entire families staying at home.
        In the months of April and June the traffic volumes of QUIC remain relatively stable, with some hours gaining traffic while other losing some.
    \item[NAT traversal / IPsec / OpenVPN:]
        Port UDP/4500 is registered at IANA for IPsec NAT traversal and is commonly used by VPN solutions, Port 
        UDP/1194 is OpenVPN's default port.
        As more people are working from home and using VPNs to access their company or university network, we see an 
        increase of both ports during working hours at the two vantage points in March.
        In the following weeks in April and June the traffic volumes for UDP/4500 stay above the traffic volume of the February base week, whereas OpenVPN's volume recedes.
        Interestingly, GRE and ESP, which transport the actual
        IPsec VPN content, decrease at the \ixpceS in March during the lockdown, while GRE traffic sees a slight increase at the \ispS.
        To summarize, more people are using VPNs from their homes resulting in an increased need of NAT traversal, but VPN connections between companies which are the primary source of GRE and ESP traffic decrease over time.
        For an in-depth analysis of VPN traffic shifts, see Section~\ref{sec:vpn}.
    \item[TV streaming:] On port TCP/8200 at the \ixpceS we see, similar to QUIC, how changes in user behavior affect the traffic profile.
        This port is used by an online streaming service for Russian TV channels.
        In March, we notice traffic volumes increasing throughout the day, shifting away from an evening centric traffic profile.
        We mainly observe this at the \ixpceS as it serves a broader and more international customer base.
        Additionally, the strong increase in March is not persisting over the following months.
    \item[Cloudflare:] Port UDP/2408 is used by the CDN Cloudflare for their load balancer service \cite{cflb}.
        We verify that the traffic indeed originates from Cloudflare prefixes.
        During our observation period, we see an increase in Cloudflare load balancer traffic at the \ixpceS in March and in June.
    \item[Video conferencing:] The video communication tool Skype and the online collaboration service Microsoft Teams 
    both use port UDP/3480, most likely for STUN purposes \cite{skypeports,teamsports}.
        We confirm this by verifying that the addresses reside in prefixes owned by Microsoft.
        Additionally, we find a small number of non-Microsoft addresses in our data.
        During the lockdown in March, we see a large increase in UDP/3480 traffic at the \ixpceS, especially during working hours on workdays.
        At the \ispS it does not show up among the top 12 transport layer ports.
         Zoom, another video conferencing solution, uses UDP/8801 for its on-premise connector which companies can deploy to route all meeting traffic through it~\cite{zoom-port-doc}.
         At the \ispS this traffic increases by an order of magnitude from February to April.
         Since Zoom only became popular in Europe due to the lockdown, this drastic increase reflects 
         the adoption of a new application by companies deploying connectors in their local network.
         These changes once again underline the fact that people working from home do change the Internet's traffic profile.
         Zoom traffic decreases again in June, which might also be related to the vacation period resulting in fewer online office meetings.
    \item[Email:] At the \ispS, especially during working hours, we find a 60\% increase in TCP/993, which is used by
    IMAP over TLS to retrieve emails. While the overall amount of traffic is small compared to, \eg QUIC, it is 
    nevertheless an additional indicator for people conducting their usual office communication from their homes.   
    \item[Unknown port:] We could not map TCP/25461 to any known protocol or service. The addresses using this port mostly reside in prefixes owned by hosting companies.
\end{description}

To summarize, we find significant changes in the traffic profile for some
popular transport-layer ports at both vantage points. This 
highlights the impact of drastic human behavior changes on traffic 
distribution during these weeks. We see an increase in work-related as well 
as entertainment-related traffic, reflecting the lockdown where people had to 
work and educate from home. This rationale 
is supported by the significant shift in workday patterns, especially at the \ispS from February to March when the 
lockdown began. As more people stay at home, the traffic levels which are 
dominated by residential customers increase steeply in the morning, compared to
the steady growth observed over the whole day in February.
\vspace{-2mm}
\section{Application classes}\label{sec:user-applicationclass}

Building on the analysis of the raw ports presented in the previous section, we now provide a more in-depth analysis of traffic shifts for different \emph{application classes}.
This is especially relevant for traffic using protocols such as HTTP(S), where a
single transport-layer port number hides many different applications and use cases.

To investigate application layer traffic shifts, we apply a traffic classification based on a combination of transport port and traffic source/sink criteria.
In total, we define more than 50 combinations of transport port and AS criteria
based on scientific-related work 
\cite{bottger2018looking, port-based-traffic-classification}, product and service documentations \cite{lol-port-information, webex-port-information, teamsports, skypeports}, and public databases \cite{ripe-query, peeringdb}.

We aggregate the filtered data into 8 meaningful application classes
representing applications consumed by end-users on a daily basis (See
Table~\ref{tab:classificationixpisp}): 
\emph{Web conferencing and telephony (Web conf)} covers all major conferencing and telephony providers, 
 \emph{Collaborative working} captures online collaboration applications, \emph{Email} quantifies email 
 communication, \emph{Video on Demand (VoD)} covers major video streaming services,
\emph{Gaming} captures traffic from major gaming providers (cloud and multiplayer), \emph{Social media} captures 
traffic of the most relevant social networks, \emph{Educational} focuses on traffic from educational networks, and 
\emph{Content Delivery Networks (CDN)} classifies content delivery traffic. Note that social networks, \eg Facebook, 
also offer video telephony and content delivery services for their own
products, which may be captured by this class but not by the more specific other classes.

\begin{figure}[t]
  \begin{subfigure}[h]{\linewidth}
    \centering
    \includegraphics[width=\linewidth]{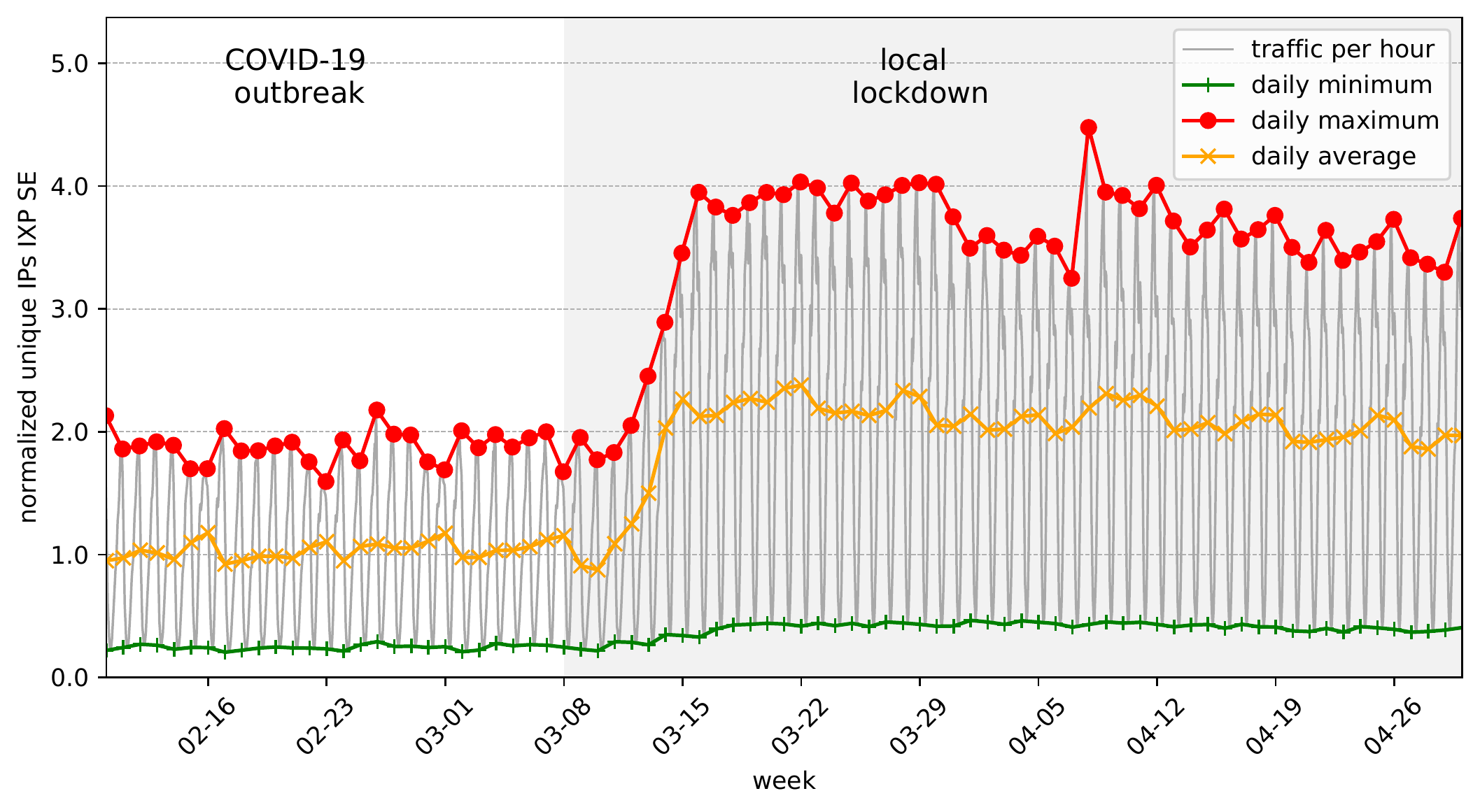}
    \label{fig:various-gaming-by-users}
  \end{subfigure}\\
  \vspace{-1.1cm}
  \begin{subfigure}[h]{\linewidth}  
    \centering 
    \includegraphics[width=\linewidth]{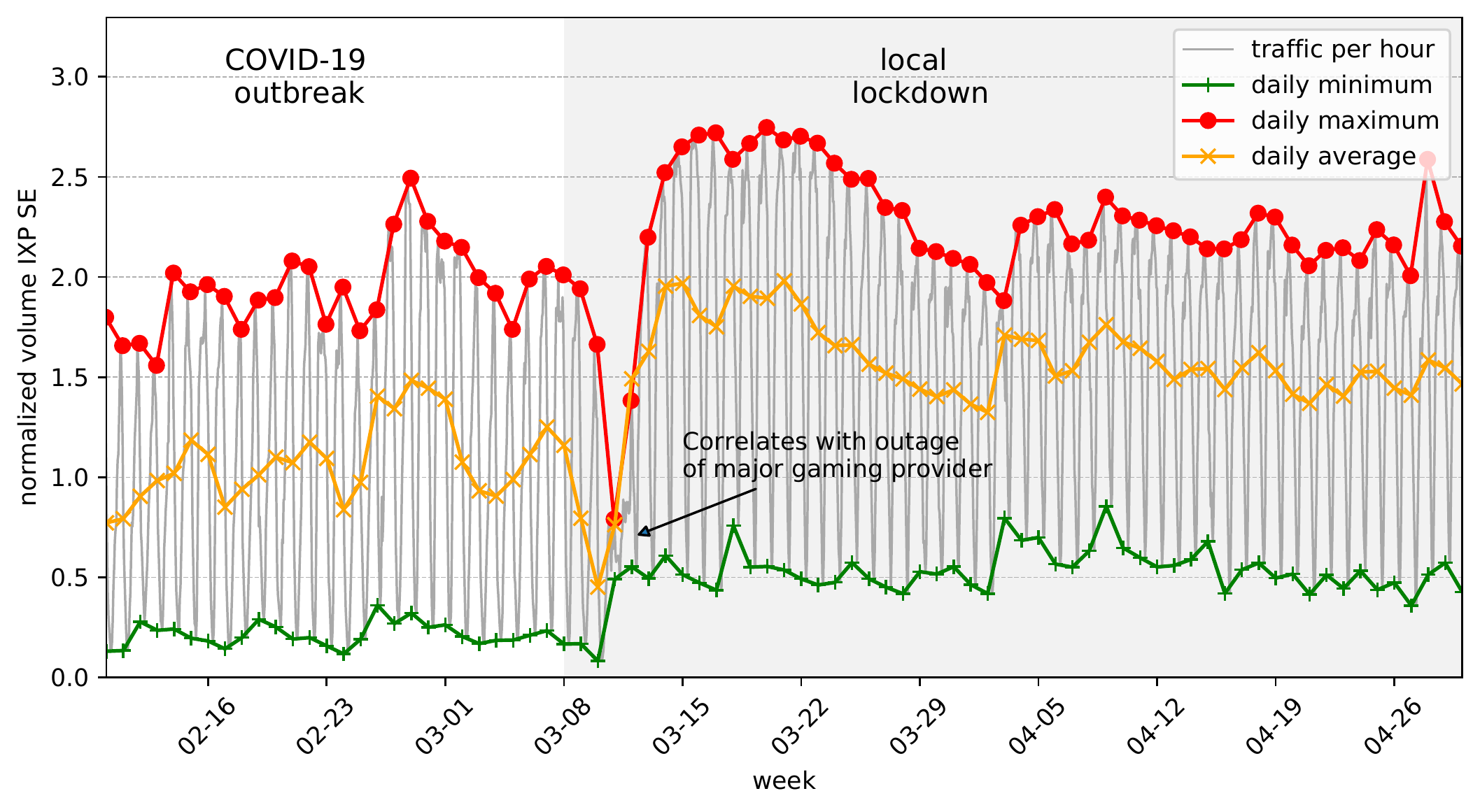} 
    \label{fig:various-gaming-by-volume}
  \end{subfigure}
  \vspace*{-2em}
  \caption{\ixpseS: Application class Gaming before and during lockdown. It
    shows a steep increase in \# IPs and traffic volume.}
  \label{fig:various-gaming}
\end{figure}

Figure~\ref{fig:various-gaming} showcases the \emph{Gaming} class at the \ixpseS vantage point.
For this application class, we filter data of five gaming software/services providers and 57 typical gaming 
transport ports in various combinations (see Table~\ref{tab:classificationixpisp}).
We then analyze the changes in usage behavior using two metrics:
\one the number of distinct source IP addresses, as a way to approximate the order of households, and \two the traffic volume.
Figure~\ref{fig:various-gaming} shows clear changes when comparing multiplayer and cloud games before and during the lockdown.
From week 10 on, \ie when the local government imposed a lockdown, the number of unique IPs seen in the trace as well as the delivered volumes rose steeply with substantial gains of the daily minimum, average, and maximum.
Notably, during the first lockdown week, the accounted volume plunges for two days to the lowest values observed in that time frame.
We verified that this is not a measurement artifact. 
Instead, the drop correlates with an outage of a large gaming provider, which may be related to the sudden increase in users.

We perform the application classification for the different IXP vantage points (\ixpseS, \ixpceS, \ixpusS) and for the 
\ispS.\footnote{In case of the \ispS we analyzed upstream as well as downstream traffic. As the differences between the 
weeks manifest in both directions in a very similar fashion we only show the downstream direction. }
To clearly present the large amount of information, we transform the data as follows.

\begin{figure*}
  \includegraphics[width=0.95\linewidth]{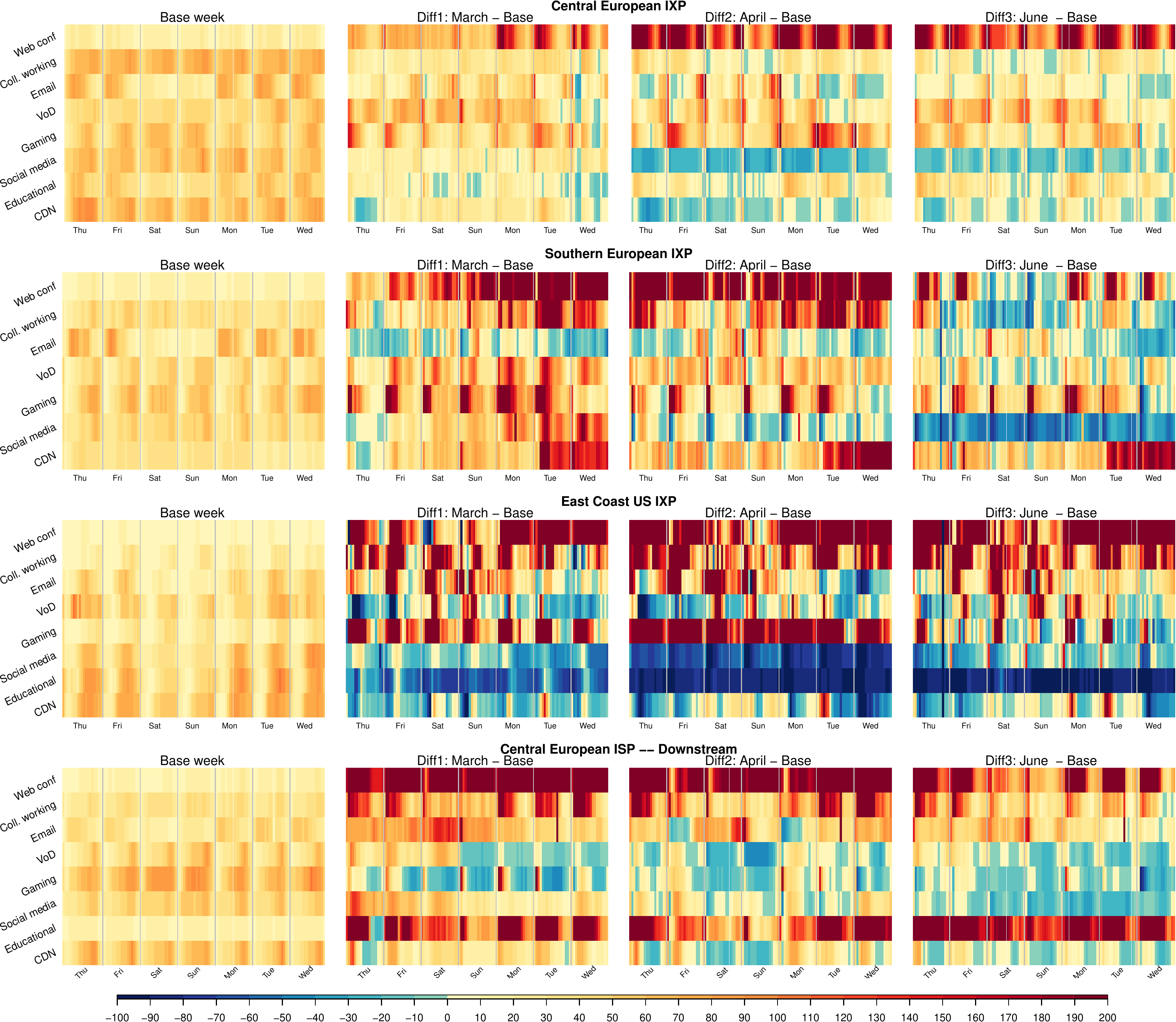} 
  \vspace*{-1em}
  \caption{Heatmaps of application class volume for three different IXP
locations and the \ispS. }
  \label{fig:ixp-heatmap-serviceclasses}
\end{figure*}

\begin{description}[leftmargin=0pt]
\item[Week-wise comparison:] We focus our analysis on four weeks, a \emph{base week} well before the lockdown, to which 
we compare three weeks representing the different stages of the COVID-19 measures as they were imposed throughout 
Europe---see Table~\ref{tab:dates} in Section~\ref{sec:datasets}.

\item[Normalization and filtering:] After normalization as outlined in Section~\ref{sec:datasets}, we remove the early morning
  hours (2--7 am).  The \emph{total} volume of the vantage points hits its daily minimum during these hours, but does not change much during
  the lockdown.  Removing these hours allows us to visualize more details of
traffic shifts during the day in order to compare
  application classes of different traffic volumes as well as the relative growth between the \emph{base week} and the other weeks.
\item[Difference to base week:] We visualize each week as the difference of the respective week and the \emph{base week}.
  This enables quick visual identification of increased/decreased application class usage compared to pre-COVID times. We remove any
  growth above 200\% and any decrease below 100\%.  
\end{description}

The condensed timelines of the different application classes are shown in Figure~\ref{fig:ixp-heatmap-serviceclasses} 
for all four vantage points. We highlight our main observations next:  

\begin{description}[leftmargin=0pt]
\item[Communication-related applications:] At all van\-tage points, \emph{Web conferencing} applications show a dramatic 
increase 
of more than 200\% during business hours, and at the \ispS, \ixpseS, and the \ixpusS also on the weekends.  In this category the \ispS experiences the largest
growth in \emph{March} right after the lockdown across all hours of the
day. In \emph{June} this trend is less pronounced, which
corresponds with people slowly going back to their offices. 
\emph{Collaborative working} mainly increases at the \ixpseS and the \ixpusS, at the \ispS we see a vast increase on 
Thursday and Friday morning which persists until \emph{June}---this might be due to coordination between work partners before the weekend. 
While in a lockdown situation one might expect a lot of additional \emph{Email}
communication, we see a different trend. At the \ixpceS and the \ixpseS
\emph{Email} actually declines during the lockdown and in \emph{June} remains on a 
lower level than before the lockdown. Instead, \emph{Email} rises at the \ispS
it, but not as high as other traffic classes as \emph{Web conferencing}. 
One possible explanation could be that many companies start connecting their remote employees via Virtual 
Private Networks (VPNs) and users connect to the mail systems via the VPN. We discuss VPN traffic in 
Section~\ref{sec:vpn}.
For the \ixpusS the trend is less pronounced, and we see phases of usage increase and decrease over time.
\item[Entertainment related applications:]\emph{VoD} streaming application usage shows high growth rates at the 
European IXPs of up to 100\%. Interestingly, \ispS only sees a slight growth of about 10\% during the lockdown, while 
in \emph{June} -- well after the lockdown -- the traffic volume drops back to the February level. Recall that the major streaming 
companies reduced their streaming resolution in Europe by mid-March~\cite{reported-date-for-netflix} for 30 days. In 
the case of the \ispS that covers the \emph{March} as well as the \emph{April} week.\footnote{The necessary measurements to 
quantify the impact of the resolution change by the VoD providers are beyond the scope of this work.} 
In the US, the trend is the other way around. Notably, this may be a biased measurement, as at the \ixpusS the 
measurement 
of the \emph{VoD} class is based on only three ASes, one of which is very large. Consequently, the decrease may reflect 
a traffic engineering decision of the large AS, e.g., establishing a private network interconnect instead of peering. 
The strong growth of \emph{gaming} applications is more coherent across all three IXP vantage points, especially 
during the day. 
While the \ispS shows a significant increase during morning hours, it generally
leans towards declining.
Note, that this effect is mainly caused by unusually high traffic levels in this category in February.
Gaming applications, typically used in the evening or at weekends, are now 
used at any time. The trend starts to flatten in June---this may in relation with people going on vacation or spending 
more time outside. 
Moreover, we see an increase at the IXPs for \emph{Social media} application traffic 
during the \emph{March} week, while the effect quickly diminishes in \emph{April}. 
In \emph{March} the ISP experiences a 70\% growth, which slows down in \emph{April} but not as drastic as at the IXPs. 
The effects in this class correlate with the gradual de-escalation of the
lockdown restrictions in Europe: as people are allowed to leave 
their homes freely again and resume social live, this traffic decreases.
In June, social media usage has returned to figures slightly below the level of March across all vantage points.
\item[Other applications:] \emph{Educational} networks and applications behave completely different at all vantage points. 
At the \ixpceS, their traffic remains relatively stable ---as would be expected given students attending 
classes from home---, but at the \ispS, instead, it drastically increases by 
up to 200\%. This growth could be attributed to some 
European educational networks providing video conferencing solutions, which are now being used by customers of the 
\ispS. Due to the lack of connected educational networks at the \ixpusS, we omit this category at this vantage point. 
See Section~\ref{sec:edu-network} for an in-depth study of the traffic shifts in
a large educational network.
Likewise, \emph{CDN} traffic increases in Europe, but does not grow much---even
decreasing at times---in the US. 
Similar to \emph{VoD}, there is a skewed distribution of CDNs present at the vantage point. Thus, a rerouting decision 
of a large player may explain the moderate loss of CDN traffic at the \ixpusS.
\end{description}

To summarize, the use of communication-related applications increase during working hours, especially in \emph{Web 
conferencing}. Entertainment related applications such as 
\emph{gaming} and \emph{VoD} are also consumed at any time of the day,
as the become more demanded during the lockdown. \emph{Social media} 
shows a strong initial increase which flattens over time. These observations
complement and strengthen those made in 
Section~\ref{sec:user-application-port}. Together, they 
demonstrate the massive impact that the drastic change in human behavior 
caused by the \covid pandemic had on application usage.

\section{VPN Traffic Shift}\label{sec:vpn}
As a response to the pandemic, many institutions asked their employees to work from home.
A typical way to access \emph{internal} company infrastructure from home is by using VPN services.
As a result, we expect VPN traffic to increase after the lockdown.

\afblock{Port-based classification.}
We apply a twofold approach to identify VPN traffic.
First, we classify traffic as VPN traffic if the well-known transport ports and protocols are used exclusively by a VPN service.
We only focus on major VPN protocols and identify IPsec (port 500, 4500), OpenVPN (1194), L2TP (1701), and PPTP (1723)---both on TCP and UDP.

\afblock{Identifying VPN usage on TCP/443.}
Since there are, however, many VPN services using TCP/443 to tunnel VPN traffic, a pure port-based identification approach cannot distinguish this traffic from HTTPS.
To \emph{limit} the potential for misclassification, we employ a second approach using DNS data to identify IPs labeled as \texttt{*vpn*} but not as \texttt{www.} in the DNS.
That is, we identify potential VPN domains by searching for \texttt{*vpn*} in any domain label \textit{left} of the public suffix~\cite{psl} (\eg \texttt{companyvpn3.example.com}) in \one 2.7B domains from TLS certificates that appeared in CT Logs during 2015--2020 and \two 1.9B domains from Rapid7 Forward DNS queries of reverse DNS, zonefiles, TLS certificates from the end of March 2020, and \three 8M domains found in the Cisco Umbrella toplist in 2020.
We resolve all matching domains to 3M candidate IP addresses.
In order to get a conservative estimate of VPN traffic over TCP/443, we then also resolve the domains from the same public suffix prepended with \texttt{www} (\eg \texttt{www.example.com}).
If the returned addresses of the \texttt{*vpn*} domain and the \texttt{www} domain match, we eliminate them from our candidates.
This approach \emph{limits} misclassifying Web traffic destined to the \texttt{www} domain as VPN traffic to the \texttt{*vpn*} domain, if they share the same IP address.
After removing shared IP addresses, we end up with 1.7M candidate VPN IP addresses.
We classify TCP/443 traffic to these VPN addresses as VPN traffic.

\begin{figure*}[tb!]
    \centering
    \includegraphics[width=\linewidth]{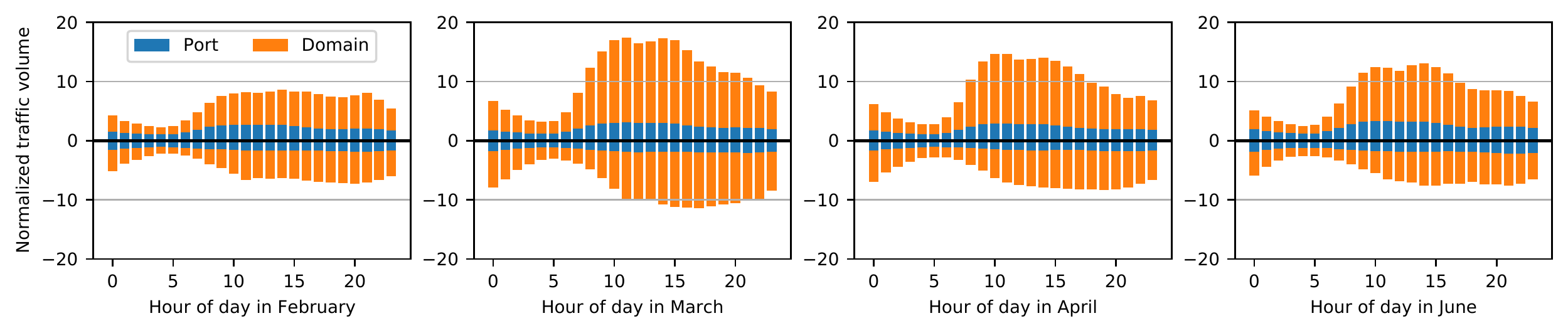}
    \caption{VPN traffic at the \ixpceS: normalized aggregated traffic volume per hour at the \ixpceS vantage point for four selected weeks. Aggregated workdays are shown as positive values, aggregated weekends as negative values. VPN servers are identified by ports and \texttt{*vpn*} label in the domain name.}
    \label{fig:vpn-ixp}
\end{figure*}

\afblock{VPN traffic on the rise.}
In Figure~\ref{fig:vpn-ixp} we report our findings using the port-based and domain-based VPN traffic identification approach.
We use four weeks of flow data from the \ixpce and aggregate them into workdays and weekends.
Interestingly, we see almost no change in port-based VPN traffic before and after the lockdown.
When looking at the VPN traffic identified with the domain-based technique, we see a significant increase in VPN traffic.
During workdays at working hours, VPN traffic increases by more than 200\% in March compared to the base week in February.
The increase on weekends is not as pronounced as during workdays, further indicating that these traffic shifts occur due to changes in user behavior (\ie people working from home).
When looking at the week in April, we still see a gain in VPN traffic compared to February, although not as large as in March.
In June, VPN traffic decreases further compared to previous months, although its
traffic volume on workdays remains well above the levels observed for the base week of February.
This is likely due to the gradual lifting of lockdown restrictions in Central Europe and the beginning of the summer holiday season, resulting in fewer people working from home in June compared to March.

In conclusion, we see a clear pattern of VPN traffic increase during working hours due to lockdown restrictions.
Moreover, as the visible increase of VPN traffic was limited to TCP/443 on \texttt{*vpn*} domains, we argue that VPN identification solely on a transport port basis vastly undercounts actual VPN traffic.
To mitigate this problem, we propose to identify seemingly HTTPS flows as VPN traffic using domain data.
This allows for a more accurate picture of the VPN landscape.

\section{Educational Network}
\label{sec:edu-network}

In this section, we study the drastic changes induced by the lockdown measures
as seen by a large European educational and research network, which
connects 16 independent universities and research centers in the metropolitan 
region of Madrid.

As a response to the pandemic, on March 9, 2020
the regional government announced the closure of the 
entire educational system from March 11 onward. Consequently, users of 
this network (\eg students, faculty, researchers, staff)
were forced to adjust and continue their teaching and research activities from 
home. Only staff for critical maintenance tasks and
security were allowed to be on-premises. 
Soon after, on March 13, the National Government
declared the national state of emergency, which was effective the next day. 
This drastic change in the activities performed at every institution
caused traffic shifts that differ noticeably from those observed in any other
vantage point we studied before. 

\afblock{Traffic volume analysis.}
We study the impact of the lockdown measures on traffic volumes at the academic
network 
by comparing three key weeks: \one one week before
announcing that the research and educational system will be closed down
(February 27 to March 4), serving as baseline,
\two the week when the lockdown happened (March 12--18) to observe the
transitioning effect, and \three a week one month after lectures moved to a
fully online model for most universities (April 16--22).%
\footnote{As opposed to the previous sections, 
we did not include the results of our traffic analysis of the
June week. At this time, Madrid was still in an intermediary stage
of the de-scalation process. Overall, EDU traffic dropped dramatically from
mid-June as most lectures and academic activities ended for the semester. 
Unfortunately, we lack access to historical traffic captures in summer time
to quantify the impact that the confinement measures had in academic traffic.}

\begin{figure}[t!]
\centering
\begin{subfigure}[t]{1\columnwidth}
\includegraphics[width=0.98\columnwidth]{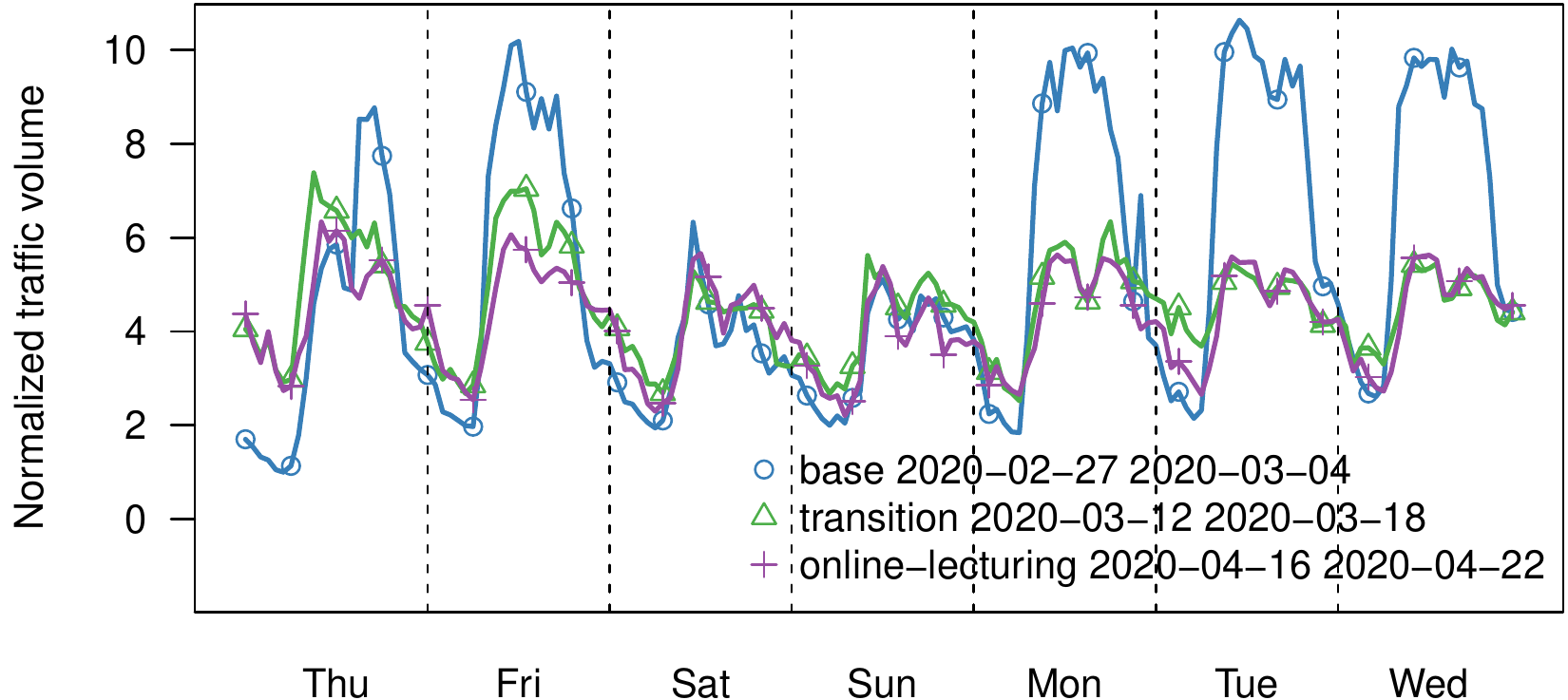}
\caption{Normalized traffic volume}
\label{fig:redimadrid-traffic-volume}
\end{subfigure}
\begin{subfigure}[t]{1\columnwidth}
\includegraphics[width=0.98\columnwidth]{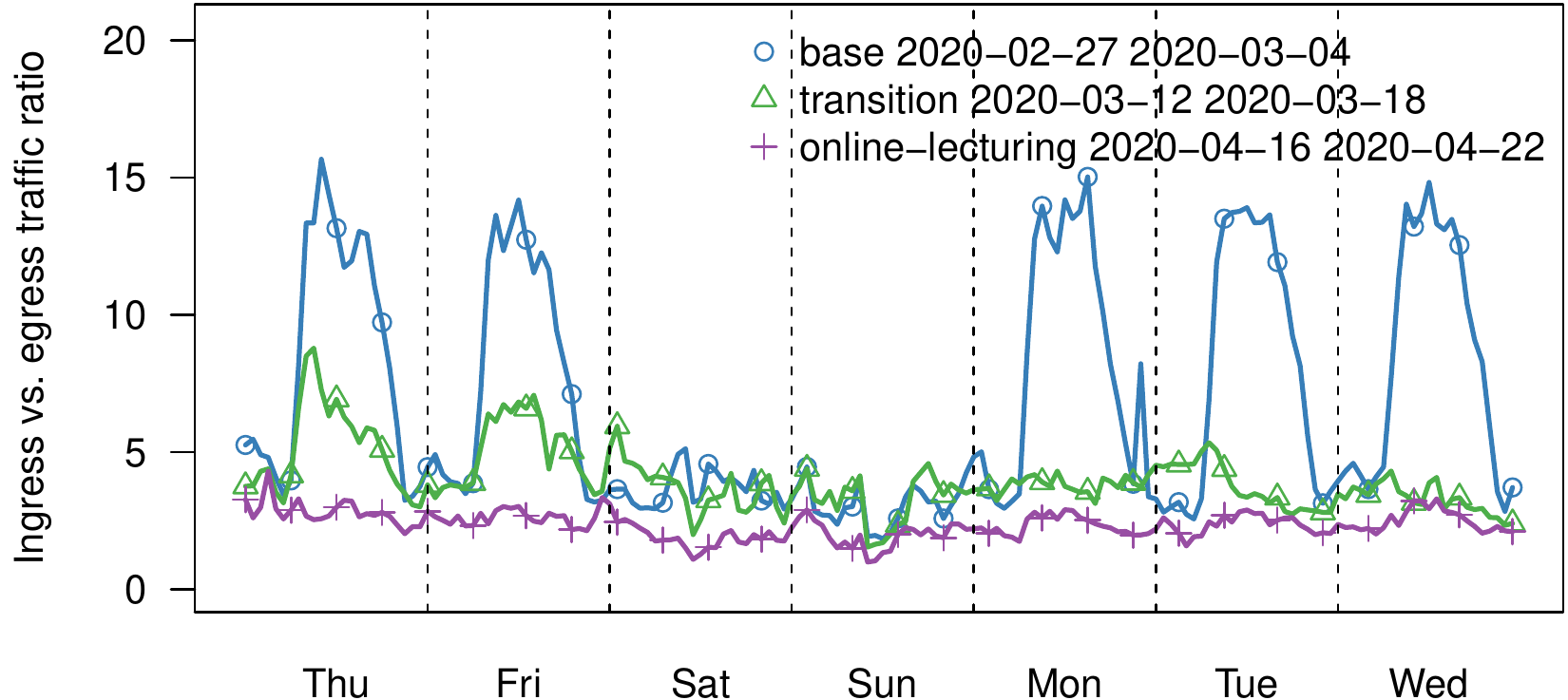}
\caption{Ingress vs.\ egress traffic ratio}
\label{fig:redimadrid-traffic-in_vs_out}
\end{subfigure}
\vspace{-0.3cm}
	\caption{\uniS: Traffic volume \& ratio \one before, \two just after,
          and \three well after the lockdown.} 
\end{figure}

Figure~\ref{fig:redimadrid-traffic-volume} shows the normalized total traffic volume
for the three weeks considered. We observe a significant drop in traffic volume
on working days between the baseline week and the two other weeks, with a
maximum \emph{decrease} of up to 55\% on Tuesday and Wednesday. Traffic on
weekends, however, \emph{increased} slightly: 14\% and 4\% on Saturday and Sunday,
respectively. The traffic reduction on working days is expected since users 
no longer use the academic network on campuses and in research facilities.  
We again observe that work and weekend days are becoming more similar
in terms of total traffic. This can be the result of a new weekly working
pattern with less distinction between both types of days due to
lockdown. Similarly, a close inspection of the hourly traffic pattern
reveals a traffic \emph{increase} from 11\% to 24\% between 9 pm and 7 am. This
could be due to users working more frequently at unusual times, but also
potentially caused by overseas students (mainly from Latin America and East
Asia as suggested by the AS numbers from which these connections come from) 
who access these resources from their home countries.

\afblock{Traffic in/out ratio analysis.}
We depict the ingress vs.\ egress traffic ratio in Figure~\ref{fig:redimadrid-traffic-in_vs_out},
showing that the ratio changed substantially after the lockdown.
In the days before the lockdown, incoming traffic was up to 15x the volume of 
outgoing traffic during workdays.
During the transition phase, the ratio halves, and it is the lowest during the third week (online lecturing), where weekend vs.\ workday pattern is no longer visible.
This change of traffic asymmetry can be explained by the nature of remote work.
On the one end, users connect to the network services mainly to
access resources, hence the increase in outgoing traffic. On the other
end, all external (\ie Internet-based) resources requested during work are no
longer accessed from the educational network but from the users' residential 
network, hence the drastic reduction in incoming traffic.

\afblock{Connection-level analysis.} 
To better understand the traffic shifts, we perform a connection-level analysis,
focusing on selected traffic classes.
We refer the reader to Appendix A %
for a list of the most relevant
classes considered in this section. We determine whether the connections are
incoming or outgoing using the AS numbers of each end-point, interfaces, and port pairs. For instance, a connection established from a residential ISP towards 
an HTTPS server hosted inside the educational network is labeled as 
\texttt{``incoming''} connection. We cannot accurately determine the 
directionality for 39\% of the flows observed at this academic network, many of 
which appear to be P2P-like applications, marginal protocols, and 
non-well-known port numbers. 

\begin{figure}[t!]
\centering
\includegraphics[width=1\columnwidth]{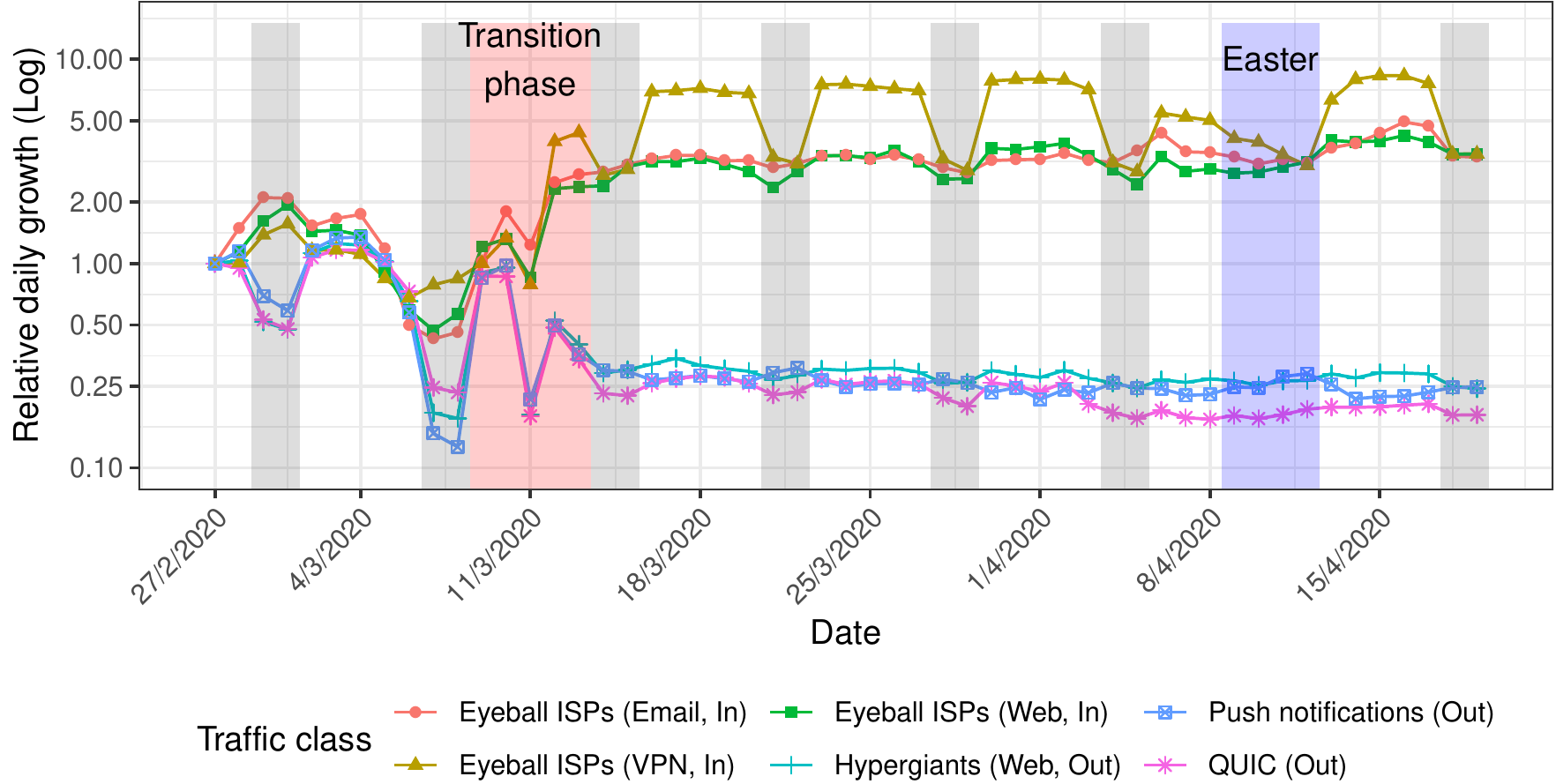}
\vspace{-0.8cm}
\caption{Daily connections relative to February 27 
for selected traffic categories. Shaded gray areas indicate weekends, red ones indicate
the transition phase enforcing confinement measures, and blue indicate the Easter break. }
\label{fig:redimadrid-traffic-conngrowth}
	\vspace{-0.3cm}
\end{figure}

The median number of the total daily connections after the declaration of the state 
of emergency grows by 24\% when compared to the pre-lockdown
baseline (ratio of median daily connections before and after March 11, 2020).
The usual 
workday-weekend differences also decrease, but are still noticeable during the 
Easter break. If we look at the directionality of the connections, 
the median number of incoming connections doubles after the lockdown, 
while the number of outgoing connections decreases almost by half. This is a
direct consequence of users having to access services hosted at the academic 
networks from the outside, which validates the observations made in the 
volumetric analysis. 

Yet, the most interesting dynamics occur for specific traffic classes.
While the average number of web connections does not
change substantially, there is a radical shift in the ratio of incoming and
outgoing connections and temporal patterns due to working from home. 
Figure~\ref{fig:redimadrid-traffic-conngrowth} shows the daily relative growth for
selected traffic categories. After the transition phase, the daily
traffic patterns for many traffic classes achieve a new status quo. 
The median number of daily incoming web connections increases by over 77\% and the number of
outgoing connections decreases by more than half. As we can see in
Figure~\ref{fig:redimadrid-traffic-conngrowth}, this reduction is even more pronounced 
for outgoing web traffic towards hypergiants or QUIC. In both cases, the number of
outgoing connections is much lower than in pre-\covid weekends. 
These drops correlate with the decrease in 
outgoing connections to push notification services and mobile services for iOS and
Android (65\% decrease on average)---\ie the number of mobile devices in the
networks decreases---as well as towards streaming services like
Spotify (83\% decrease).

We focus on those traffic classes that are associated with remote
working and lecturing. Table~\ref{tab:classificationedu} provides 
the definition of the classes discussed in this section. Precisely, 
the median incoming connection growth for web, email, VPN, Remote Desktop, and 
SSH connections is 1.7x, 1.8x, 4.8x, 5.9x, and 9.1x respectively. The significant
increase in incoming web traffic is caused by users accessing online 
teaching material and other resources hosted at some of these universities,
primarily from eyeball ISPs from the same country (2.8x growth). 
As mentioned in the volumetric analysis, we can observe a shift in the 
hourly connection patterns for both incoming and outgoing web connections. 
Traditional working hours are still noticeable---including a drop in connections
during lunch---but after the \covid outbreak, 
a significant fraction of users access these services 
late in the evening as well as early in the morning. If we analyze the origin
ASes for these out-of-time  
connections, we can observe that many connections are established from overseas 
eyeball ASes from Latin America (1.8x), but also from North America (3.4x). 
In fact, time zone differences are noticeable. National users access
web resources hosted at the university from 10 am to 9 pm, with a valley from 2
to 4 pm. Latin American users start connecting at 5 pm,
presenting a peak from midnight until 7 am (peak hours are 3 and 4 am).
Interestingly, while the temporal patterns for VPN, web, and remote desktop are
correlated, SSH traffic patterns are irregular.

\afblock{Takeaway.} Academic networks experience drastic traffic
shifts due to \covid. Traffic volume, directionality ratios, and its source and 
destination are radically different from before \covid.
This behavior is antagonistic, yet complementary, to the one observed 
in residential ISPs.
%

%
 %
\vspace{-2mm}
\section{Related Work}\label{sec:related-work}

Our study provides a testimonial of the impact of an unprecedented medical crisis in recent human history on the operation of the
Internet. Previous studies followed a similar approach to ours, \ie collect measurements at different vantage
points, to understand the impact of other events on the Internet. Partridge \etal collected and
analyzed routing and protocol data during and after 9/11 to understand the resilience
of the Internet under stress~\cite{NAP-network-911}. Their findings showed that, overall, the
Internet operation was robust: Although
unexpected outages did happen, they only had a local impact. Notice, however, 
that the penetration and importance of the
Internet in our life has significantly increased in the last twenty years, and
the global nature of the \covid pandemic crisis makes this case unique. 
Other studies focus on physical phenomena,
\eg earthquakes~\cite{Earthquake-Japan} or severe weather
conditions~\cite{weather-sigcomm2019,ZMap}, 
and power outages~\cite{down-the-failure,bogutz2019identifying} to understand the Internet behavior and the change on
Internet user activity.
Beyond physical phenomena, also human-triggered events such as major update roll-outs can cause substantial traffic shifts~\cite{applecdn}.

The study of the impact of the \covid pandemic to the performance and traffic of the Internet has attracted
significant attention in the form of
blogs posts~\cite{Akamai-30pcUS-pandemic,Akamai-30pcEU-pandemic,Comcast-covid19,Telegeography-covid19,Google-covid19}
and more recently in presentations at network operator
conferences~\cite{Labovitz-covid-19}. By the time of our submission, a limited
number of research studies have been already
published. Favale \etal report and analyze the impact of the remote learning
activity by 16k students on the Politecnico di Torino campus network due to the lockdown enforcement \cite{covid19-Polito-campus-traffic}. The university
utilized an in-house online teaching solution.
Thus, although the impact of remote learning on the campus network shares similarities with our analysis of the
academic and research network in our study, there are also significant differences.  Another study~\cite{covid19-wifi-campus-traffic}
analyzed Wi-Fi network data collected at university campuses in Singapore and the US during the pandemic.  
Their results show that the activity on campuses decreases, but mobility did not, 
as this would require more drastic measures by the governments. In our study, we found that the mobility patterns
reduced drastically in Europe, most likely due to the stricter measures and complete lockdowns. A study of the access
patterns of Wikipedia shows that during the pandemic Web visitors had an
increased interest in topics such as health~\cite{covid19-wikipedia}. 
This increase was even more pronounced for countries that were in the epicenter of the
pandemic.
Parallel to our work, researchers evaluated \one the impact of the pandemic on traffic of a UK mobile network operator reflecting changes in users' mobility~\cite{CovidTelefonica}, \two changes in traffic demand at a major social network~\cite{FacebookCovid}, \three transactions volumes at an underground market during the pandemic concluding that the observed higher transaction volumes are a market stimulus rather than an effect of the pandemic~\cite{CovidCybercrime}, and \four the impact of the pandemic on Internet latency in various European countries, finding an increase in the variance of additional latency and packet loss \cite{candela2020impact}.
\section{Discussion}\label{sec:discussion}

\afblock{Internet operation during the pandemic: a success story.}
The \covid pandemic ``underscored humanity's growing reliance on digital networks for business continuity, employment,
education, commerce, banking, healthcare, and a whole host of other essential
services''~\cite{ITU-press-release-COVID}. At the beginning of the pandemic,
changes in user demand for online services raised concerns for network
operators, e.g., to keep networks running smoothly especially for life-critical organizations such as
hospitals~\cite{snijdersNetOps}. 
In fact, the pandemic increased the demand for applications supporting remote
teaching and working to guarantee social distancing as shown in our
analysis across all vantage points.
The Internet could handle this new load due to the flexibility and elasticity that cloud services
offer, and the increasing
connectivity of cloud providers~\cite{Arbor:SIGCOMM2010,Labovitz-2009-2019,one-hop-Inet,Edge-Fabric-2017,Espresso-2017}. Our results confirm that most of the applications with
the highest absolute and relative increases are cloud-based.
Moreover, the adoption of best practices on designing, operating, and provisioning networks contributed to the smooth
transition to the new normal. Due to the advances in network automation and deployment, e.g., automated configuration management and robots installing
cross connects at IXPs without human involvement, it was possible to cope with the increased demand.
For example, DE-CIX Dubai managed to quickly enable new ports within a
week for Microsoft which was selected as the country's remote teaching solution for high
schools~\cite{decixDubaiHomeschooling}. 
In summary, our study demonstrates that 
over-provisioning, network management, and automation are key to provide
resilient networks that can sustain drastic and unexpected 
shifts in demand such as those experienced during the \covid pandemic.

\afblock{Taming the traffic increase.} In this paper, we report an increase in
traffic in the order of 15-20\% within days after the lockdown began. This is in line
with reports of ISPs and
CDNs~\cite{Labovitz-covid-19,Comcast-covid19,Akamai-30pcUS-pandemic,Akamai-30pcEU-pandemic}
as well as IXPs~\cite{ripeIXPs}. 
Typically, ISPs and CDNs are prepared for a traffic increase of 30\% in a single year
period~\cite{Labovitz-2009-2019,Akamai-Inet-works-pandemic,cisco-anrep}.
While these are yearly plannings, the pandemic created substantial shifts within
only a few days.
As a result, ISPs either needed to benefit from over-provisioned capacity---e.g., to handle unexpected traffic spikes such as attacks or flash-crowd events---or
add capacity very quickly.
We observed port capacity increases in the order of 1,500 Gbps (3\%) across many IXP members
at the \ixpceS alone (see Section~\ref{sec:macroperspective}).
Beyond our datasets, some networks publicly reported that traffic shifts due to
the pandemic resulted in partial connectivity issues and required new interconnections~\cite{dfnCovid, ripe80NewPoP}. %
When we turn our attention to traffic peaks, we notice
that the increase is even smaller. Traffic engineering focuses on peak traffic increase as this requires more network
resources. The effect of the pandemic fills the valleys during the working hours and has a
moderate increase in the peak traffic, which can be handled by well-provisioned networks that are prepared for
sudden surges of peak traffic by 30\% or more, due to attacks, flash-crowds, and link failures that shift traffic to
other links.
One concern that network operators raised in March brought awareness to network instabilities that might occur due to traffic shifts~\cite{snijdersNetOps}.
While on the one hand we find no evidence that the traffic shifts due to the pandemic
  impact network operation of our vantage points, individual links 
  experience drastic increases in traffic---way beyond the overall 15-20\%. 
Such increases arise unexpectedly to some network operators and may create a
need for port upgrades. On the other hand, the vantage points in this paper range from extremely large to moderate sizes with 
sufficient resources and a lot of experience in network provisioning and resilience. In general, smaller networks with 
limited resources may not be able to plan with sufficient spare capacities and fast enough reaction times to compensate 
for such sudden changes in demand.

\afblock{Substantial shift in traffic pattern.}
From a network operator perspective, coping with the pandemic has required some port capacity upgrades but otherwise
does not appear to impact operation.  The ability of network operators to quickly add capacity when needed highlights
that the Internet infrastructure works well at large, despite some challenges to access data centers imposed by the
lockdown.  From the perspective of the traffic mix, the pandemic, however, results in substantial changes in traffic,
ranging from shifted diurnal pattern to traffic composition.  This represents a remarkable shift in Internet traffic
that is, based on our observations, handled surprisingly well by the Internet core at large supposedly because many operators are prepared and can react
quickly to new demands.  While the pandemic represents a rather extreme and exceptional case, one may argue that with
the growing intertwining of the Internet and our modern society such events can occur more often.  In any case, the
\covid pandemic highlights that user behavior can change quickly and network operators need to be prepared for sudden
demand changes. \todo{we have to revisit the next sentences:} 

\section{Conclusion}\label{sec:conclusion}

The \covid pandemic is a---hopefully once in a lifetime---event that 
drastically changed working and social habits for billions of people. 
Yet, life continued thanks to the increased
digitization and resilience of our societies, with the Internet 
playing a critical support role for businesses, education, entertainment, 
and social interactions.
In this paper, we analyzed network flow data from multiple vantage points,
including a large academic network and a large ISP at the edge,
and, at the core, three IXPs located in Europe and the US. Together, 
they allow us to gain a good understanding of the lockdown effect on 
Internet traffic in more developed countries.

Our study reveals the importance of using different lenses to fully 
understand the \covid pandemic's impact at the traffic level: Mornings and 
late evening hours see more traffic. Workday traffic patterns are rapidly 
changing and the relative difference to weekend patterns is disappearing. 
Applications for remote working and education, including VPN and video 
conferencing, see traffic increases beyond 200\%. For other parts of the 
Internet, such as educational networks serving university campuses, we 
find decreasing traffic demands due to the absence of users but a drastic
increase in certain applications enabling remote working and lecturing. 
For some networks, we observe that traffic ratios---including  
sources and destinations---, are radically different from a pre-\covid 
pandemic scenario. These observations highlight
the importance of approaching traffic engineering with a focus that
looks beyond hypergiant traffic and popular traffic classes to
consider ``essential'' applications for remote working.

With the evidence provided in this paper, we conclude that the 
Internet---from the perspective of our vantage points---did its job and 
coped well with unseen and rapid traffic shifts. Related work, however, 
reported performance degradation in less developed regions~\cite{FacebookCovid}. 
The unseen traffic shifts we observe due to the implementation of confinement
measures underline the importance of the Internet's distributed nature to 
react amicably to such events and enhance society's resilience.

\subsection*{Acknowledgements}

The authors would like to acknowledge the support offered by David Rinc\'{o}n
and C\'{e}sar S\'{a}nchez (IMDEA Software Institute and REDIMadrid) to access the
academic network dataset. This work has been partially funded by
the Federal Ministry of Education and Research of Germany (BMBF, grants ``Deutsches Internet-Institut'' 16DII111,
5G-INSEL 16KIS0691 and AIDOS 16KIS0975K), the Spanish Ministry of Science, Innovation and Universities (grant
TIN2016-79095-C2-2-R), and by the Comunidad de Madrid (grants EdgeData-CM P2018/TCS-4499 and CYNAMON-CM P2018/TCS-4566,
co-financed by European Structural Funds ESF and FEDER), and by the European Research Council (ERC) Starting Grant ResolutioNet (ERC-StG-679158).
The opinions, findings, and conclusions or recommendations expressed are those of the authors and do not necessarily reflect those of any of the funders.

\bibliographystyle{ACM-Reference-Format}
\bibliography{paper}

\appendix

\appendix
\section{Traffic Classification}\label{appendix:trafficclassification}

To the best of our knowledge, there is no established and comprehensive classification of flow data into traffic classes. 
Even if such a classification existed, it would be a constantly moving target and highly dependent on the vantage point.
Thus, we have compiled classifications based on scientific-related work such as \cite{bottger2018looking, port-based-traffic-classification}, product and service documentations \cite{lol-port-information, webex-port-information, teamsports, skypeports}, and public databases \cite{ripe-query, peeringdb} for the different vantage points.
These classifications have the largest possible overlap, but may differ between vantage points for one or more of the following reasons.

\afblock{Local differences.} We are investigating vantage points from a total of three countries on two continents. There exist local content providers and ISPs in each country that play a dominant role in their respective home market (e.g., digital offers of local broadcasting networks, national ISPs). Likewise, for IXPs, not every network is present at every IXP, which makes defining a common classification across different IXPs difficult.

\afblock{Different types of Networks.} We investigate different types of
networks attracting different traffic mixes. For instance, cloud gaming does not
play a major role in academic networks (see Section~\ref{sec:edu-network}), and Video on Demand is usually not consumed via mobile providers. Consequently, different traffic classes are relevant for different networks leading to a different classification.

\afblock{Ease of Classification.} Not all traffic classes can be classified
easily and they are not mutually exclusive. An example is the VPN classification in \autoref{sec:vpn} requiring the additional use of DNS information. Moreover, the number and size of the datasets used in this work is exceptional, so certain classifications cannot be performed on all data in reasonable time.

Notably, the goal of the classifications defined in this work is not to catch \emph{all} traffic for a certain traffic class, but rather a \emph{representative} subset of traffic allowing to reason about trends during the pandemic.
In the following we disclose as many details of the classifications used in this work as possible. However, due to the sensitive nature of flow data, some of the information is covered by non-disclosure agreements and can therefore not be published.

\subsection{Hypergiants Classification}
\label{appendix:hypergiants}

A classification of hypergiant ASes is provided by Böttger \etal~\cite{bottger2018looking}. As this classification is established in the scientific community, we leverage it in this work. For more details on this classification and how the 15 ASes are selected, see ~\cite{bottger2018looking}. Table~\ref{table:1} reports the full list of ASes considered for this category. Nevertheless, the classification is limited to a few very large networks and cannot provide insights beyond these hypergiants.

\begin{table}[t]
\footnotesize
\centering
\begin{tabular}{c c } 
 \hline \toprule
\textbf{Org. Name}	& \textbf{ASN} \\ \midrule
Apple Inc & 714 \\
Amazon.com & 16509 \\
Facebook & 32934 \\
Google Inc. & 15169 \\
Akamai Technologies & 20940 \\
Yahoo! & 10310 \\
Netflix & 2906 \\
Hurricane Electric & 6939 \\
OVH & 16276 \\
Limelight Networks Global & 22822 \\
Microsoft & 8075 \\
Twitter, Inc. & 13414 \\
Twitch & 46489 \\
Cloudflare & 13335 \\
Verizon Digital Media Services & 15133 \\
 \hline
\end{tabular}
\caption{List of Hypergiant ASes as defined by Böttger \etal~\cite{bottger2018looking}. Used to classify data in 
Figures~\ref{fig:hypergiant-other} and \ref{fig:redimadrid-traffic-conngrowth}.}
\label{table:1}
\vspace{-2em}
\end{table}

\subsection{Application Classification Academic Network}
\label{appendix:classificationacademic}

For the academic network, we focus on applications we expect to be used by
academic staff and students, \eg VPN, SSH, remote desktop applications and
entertainment (\eg Spotify), see Table~\ref{tab:classificationedu}.

\begin{table}[!t]
\footnotesize
\begin{tabular}{p{2cm}p{5.5cm}}
\toprule
\textbf{Application class} & \textbf{Filter} \\ \midrule
Web	& TCP:80, TCP:443, UDP:443 (QUIC), TCP:8000,
TCP:8080 \\
QUIC &  UDP:443 \\
Push notifications & TCP:5223, TCP:5228 \\
Email & TCP:25, TCP:110, TCP:143, TCP:465, TCP:587,
TCP:993, TCP:995 \\
VPN & UDP:500, ESP, GRE, TCP:1194, UDP:1194, UDP:4500
(Fortigate) \\ 
SSH & TCP:22 \\
Remote Desktop & TCP:1494 and UDP:1494 (Citrix remote desktop), 
TCP:3389 (Windows remote desktop), TCP:5938, UDP:5938 (Team Viewer) \\
Spotify & TCP:4070 or ASN8403 \\ \bottomrule
\end{tabular}
\caption{Overview of filters for the EDU traffic application classification analysis (Section~\ref{sec:edu-network}). We note that these 
categories are not mutually exclusive (\eg QUIC is a subset of Web) 
to enable the analysis of different phenomena.}
\label{tab:classificationedu}
\vspace{-2em}
\end{table}

\begin{table*}[t]
\footnotesize
\begin{tabular}{@{}lccc@{}p{9cm}}
\toprule
\textbf{Application class }             & \rot{\textbf{\# of filters}} &
\multicolumn{1}{l}{\rot{\begin{tabular}[c]{@{}l@{}}\textbf{\# of distinct} \\
\textbf{ASNs}\end{tabular}}} &
\multicolumn{1}{l}{\rot{\begin{tabular}[c]{@{}l@{}}\textbf{\# of distinct} \\
\textbf{transp. ports}\end{tabular}}} & \textbf{Notes} \\ \midrule
Web conferencing and telephony (Web conf) & 7  & 1  & 6  & Conferencing audio/video ports, AS-based for pure conferencing provider (TCP:444, UDP:3478-3481, UDP:8200, UDP:5005, UDP:1089, UDP:10000) \\
Video on Demand (VoD)          & 5             & 5  & -  & Large to medium VoD provider ASes \\
Gaming                         & 8             & 5  & 57 & Transport ports of popular games , AS-based for large gaming providers (e.g. TCP:1716, TCP:4001, TCP:3074, ...), includes cloud gaming services \\
Social media                   & 4             & 4  & 1  & Social networks including their respective CDNs (HTTPs+respective AS) \\
email                          & 1             & -  & 10 & Typical mail transport ports (TCP:25, TCP:587, TCP:109, TCP:110, TCP:143, TCP:220, TCP:645, TCP:585, TCP:993, TCP:995) \\
Educational                    & 9             & 9  & -  & ASes of universities close to respective vantage points \\
Collaborative working          & 8             & 2  & 9  & Collaborative editing, file sharing, versioning, VPN, remote administration (e.g. TCP:1194, UDP:1194, UDP:1197, UDP:1198, ...) \\
Content Delivery Network (CDN) & 8             & 8  & -  & Dominant CDN providers (excluding social network CDNs) by AS \\ \bottomrule
\end{tabular}
\caption{Overview of filters for the application classification. Filters are based on transport ports or ASes , either 
in combination or separately. Used to classify data in Figures \ref{fig:various-gaming}, 
\ref{fig:ixp-heatmap-serviceclasses}.}
\label{tab:classificationixpisp}
\vspace{-1em}
\end{table*}

\subsection{Application Classification ISP/IXPs}
\label{appendix:applicationclassificationispsixps}

As ISP and IXP networks have a comparable traffic mix, we compiled a joint classification for the ISP/IXP vantage points
allowing for a high comparability. The classification is based on combinations of ASes (at IXPs by port, at ISP by IP
ranges) and transport protocol ports if characteristic protocols exist. While the transport protocols are disclosed in
\autoref{tab:classificationixpisp}, the measured ASes cannot be disclosed due to non-disclosure agreements.

\section{Additional Plots for Link Utilization}
\label{appendix:link-utilization}

\begin{figure}[t]
	\centering
	\includegraphics[width=\columnwidth]{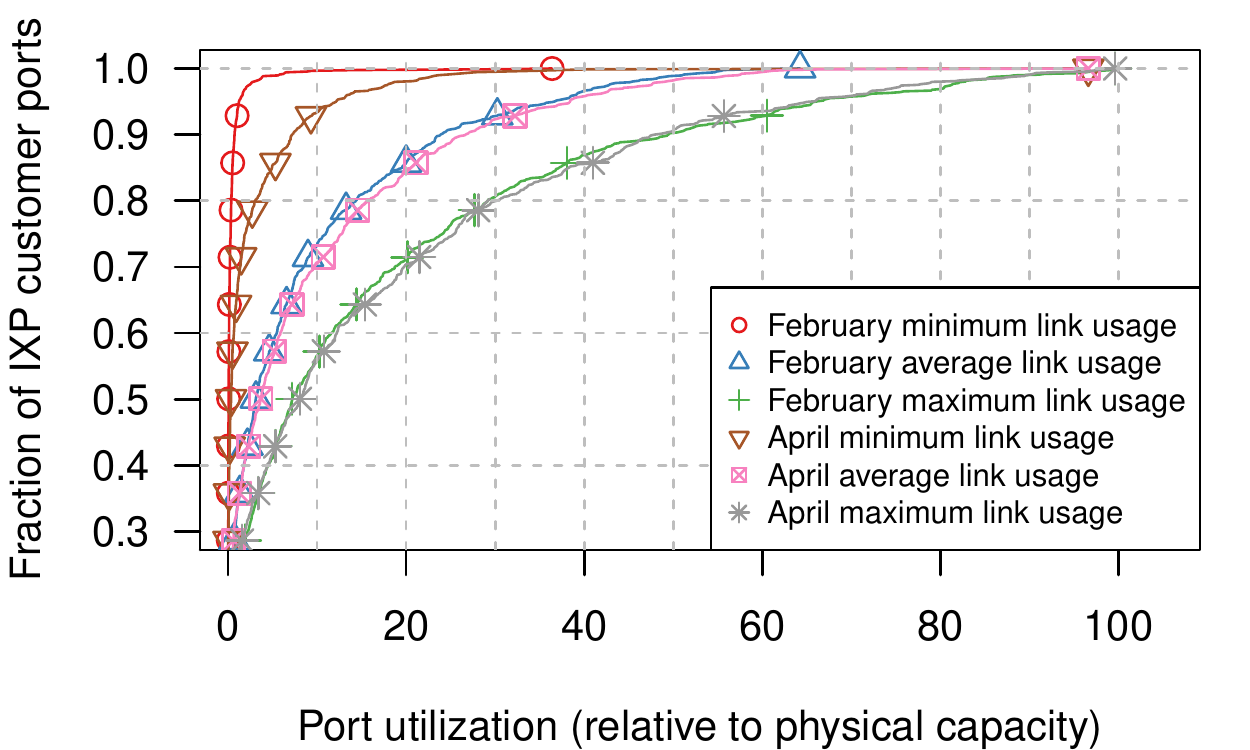}
	\caption{ECDF of minimum, average and maximum link utilization at \ixpceS, February week vs. April week.}
	\label{fig:link-utilization-april}
\end{figure}

\begin{figure}[t]
	\centering
	\includegraphics[width=\columnwidth]{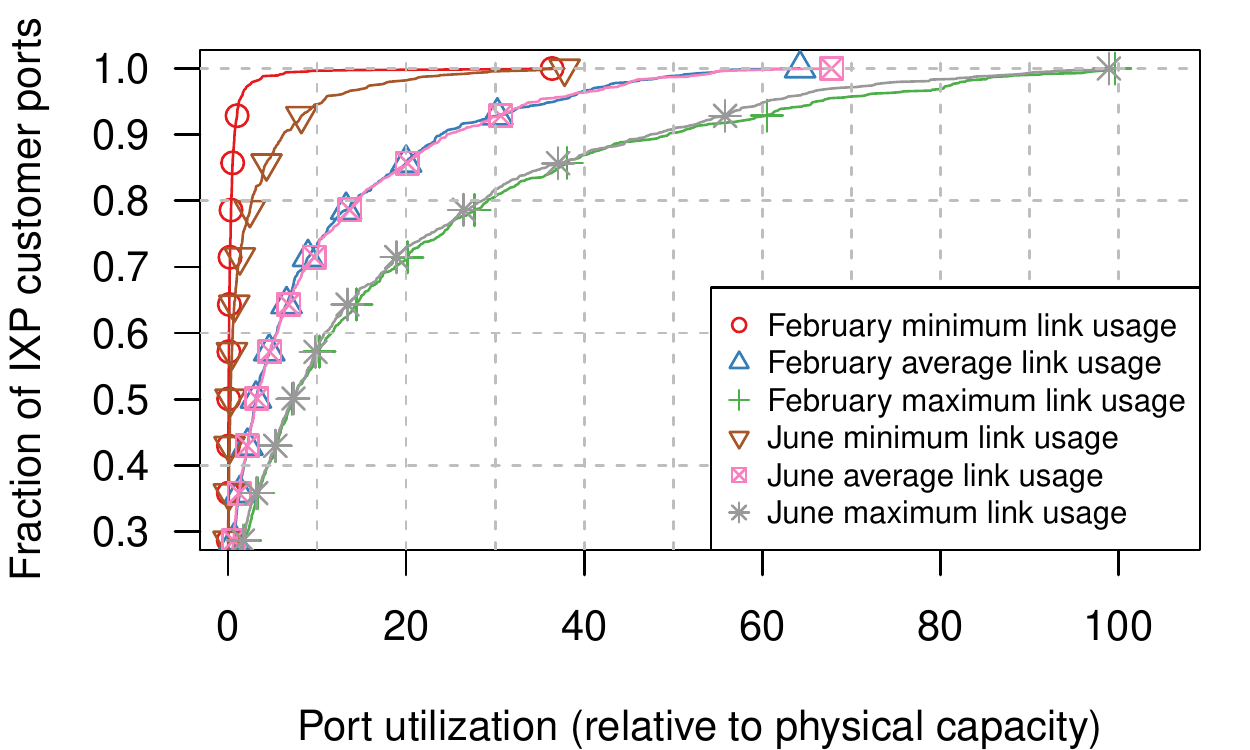}
	\caption{ECDF of minimum, average and maximum link utilization at \ixpceS, February week to June week.}
	\label{fig:link-utilization-june}
\end{figure}

The hereby presented plots serve as an addition to statements made in Section \ref{subsec:link-utilization}.
Figures \ref{fig:link-utilization-april} and \ref{fig:link-utilization-june} 
show the relative link utilization at \ixpceS for weeks in April and June, respectively. We also plot the link utilization from the reference week in February for
comparison.  These plots show, in contrast to Figure \ref{fig:shift-link-utilization-ixp}, an increased overall link
utilization at \ixpceS.

\begin{figure*}[t]
    \centering
    \includegraphics[width=\textwidth]{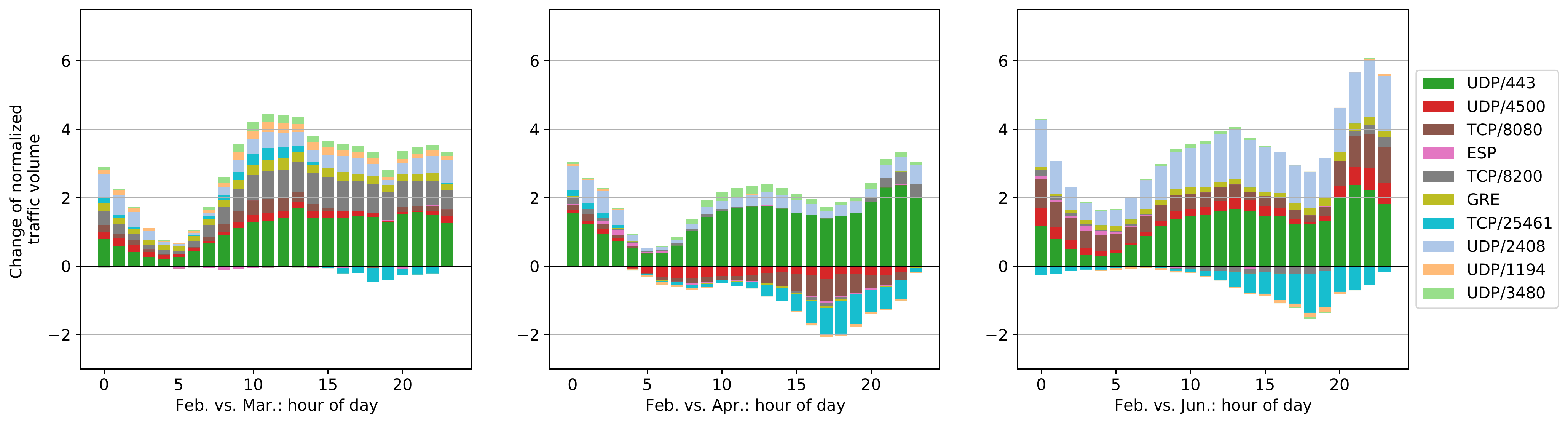}
    \caption{\ixpceS traffic difference by top application ports: normalized aggregated
        traffic volume difference per hour comparing the weekends of March, April, and June to the base week of February.
        We omit TCP/80 and TCP/443 traffic for readability purposes.}
    \label{fig:ports-ixp-first}
\vspace{-1em}
\end{figure*}

\begin{figure*}[t]
    \centering
    \includegraphics[width=\textwidth]{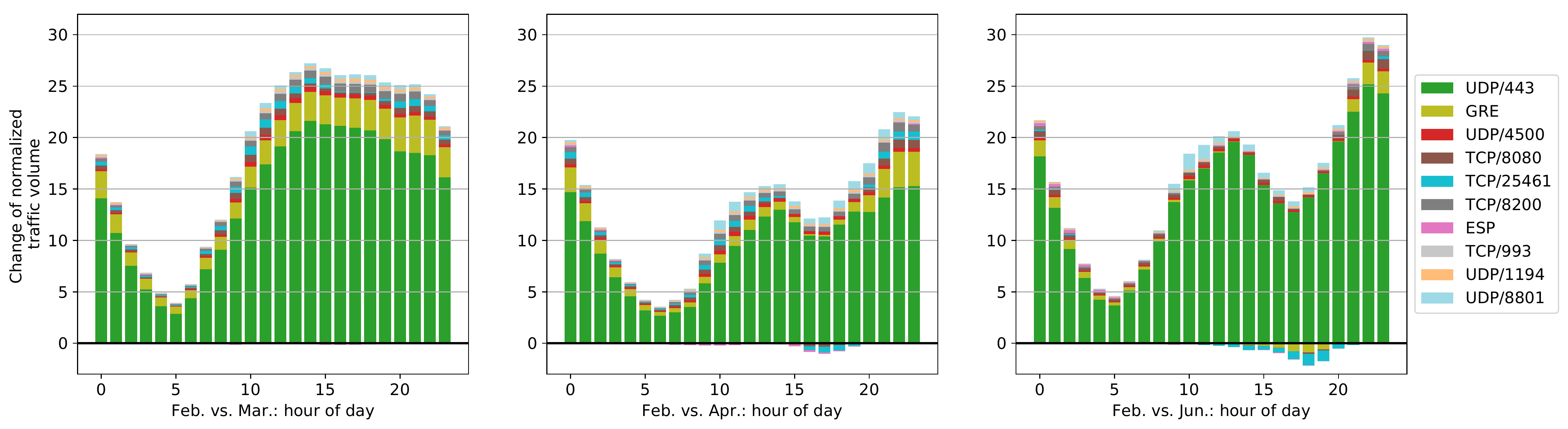}
    \caption{\ispS traffic difference by top application ports: normalized aggregated
        traffic volume difference per hour comparing the weekends of March, April, and June to the base week of February.
        We omit TCP/80 and TCP/443 traffic for readability purposes.}
    \label{fig:ports-isp-first}
\vspace{-1em}
\end{figure*}

\begin{figure*}[t]
    \centering
    \includegraphics[width=\textwidth]{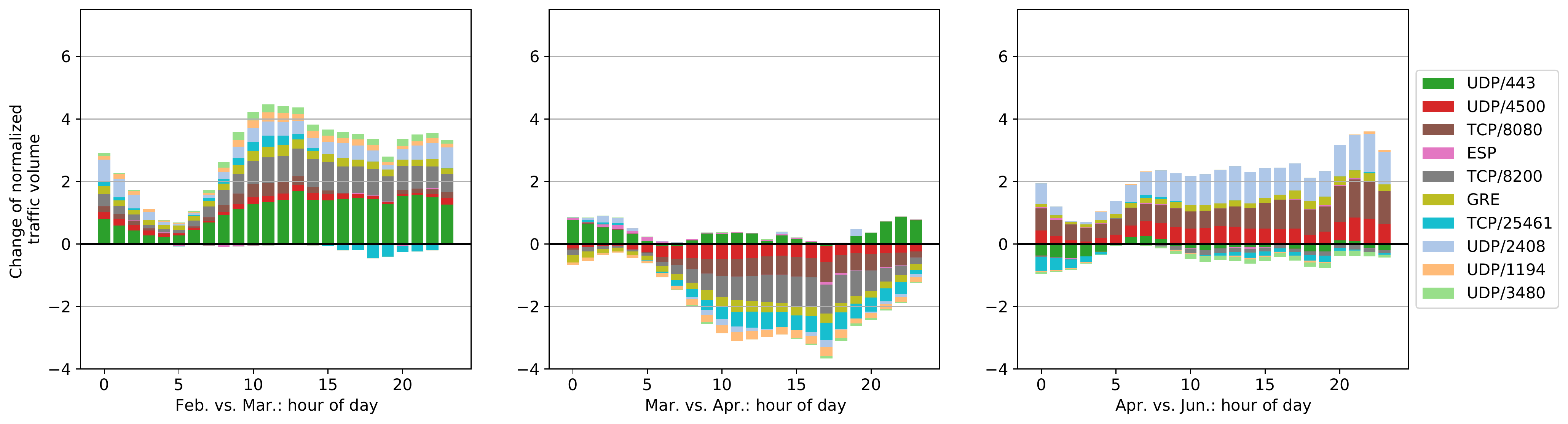}
    \caption{\ixpceS traffic difference by top application ports: normalized aggregated 
        traffic volume difference per hour comparing the weekends of February, March, April, and June.
        We omit TCP/80 and TCP/443 traffic for readability purposes.}
    \label{fig:ports-ixp-weekend}
\end{figure*}

\begin{figure*}[t]
    \centering
    \includegraphics[width=\textwidth]{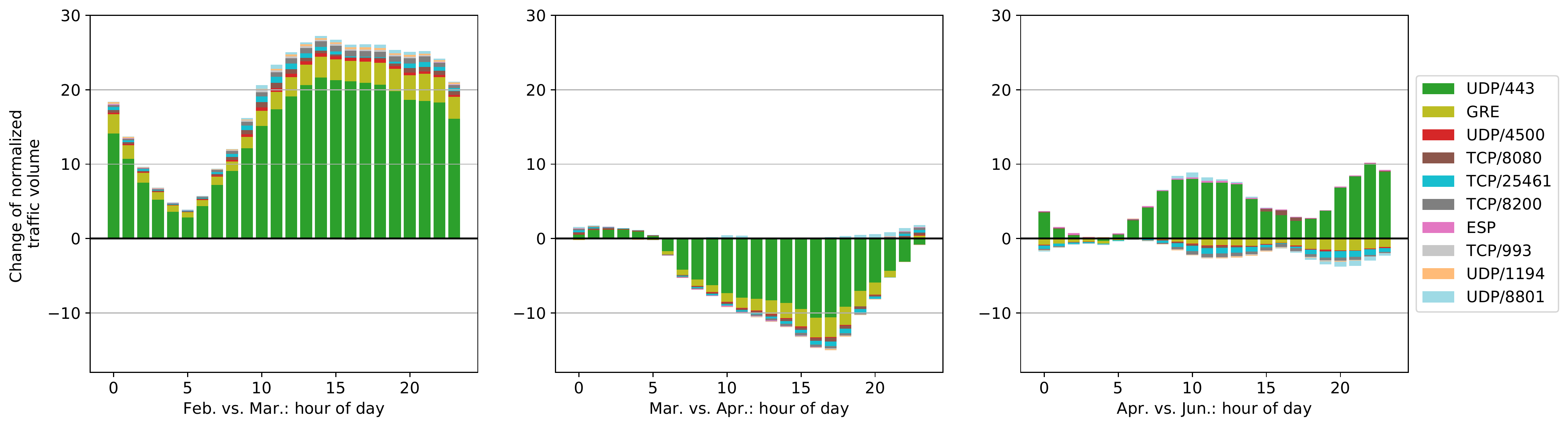}
    \caption{\ispS traffic difference by top application ports: normalized aggregated
        traffic volume difference per hour comparing the weekends of February, March, April, and June.
        We omit TCP/80 and TCP/443 traffic for readability purposes.}
    \label{fig:ports-isp-weekend}
\end{figure*}

\section{Additional Plots for Applications by Port Classification}
\label{appendix:application-port}

In the following we present additional plots for the applications by port classification.

Figures~\ref{fig:ports-ixp-first} and \ref{fig:ports-isp-first} show the differences by top application ports compared directly with the base week of February, for the \ixpceS and \ispS respectively.
This is different from Figures~\ref{fig:ports-ixp} and \ref{fig:ports-isp} shown in Section~\ref{sec:user-application-port} which show the difference for weeks of two consecutive month, \ie emphasizing on the differences between the selected weeks.
On the other hand, Figures~\ref{fig:ports-ixp-first} and \ref{fig:ports-isp-first} emphasize the result of the lockdown and its lifting compared to the regular February 2020 week, \ie what changes do we observe in each month compared to the base week.

All previously shown figures focusing on the applications by port classification, are limited to the changes for
workdays within the selected weeks.  To complement the changes seen on workdays, Figures~\ref{fig:ports-ixp-weekend} and
\ref{fig:ports-isp-weekend} show the traffic changes based on application port and hour of weekend days only.

\end{document}